\documentclass[aps,preprint]{revtex4}
\usepackage{amsmath,amssymb,amsthm,mathrsfs,amsfonts,dsfont} 
\usepackage{graphicx}
\usepackage{subfigure}
\usepackage{amsbsy}
\usepackage{bm}
\usepackage{xcolor}
\usepackage{dcolumn}
\newcommand{\blue}[1]{{\color{black}#1}}
\def\bx{\mathbf{x}}
\def\id{\mathds{1}}
\def\a{\mathbf{a}}
\def\J{\mathbf{J}}
\def\j{\boldsymbol{\xi}}
\usepackage{algorithm}
\usepackage{algpseudocode}
\usepackage{amsmath}
\usepackage{array}
\usepackage{booktabs}

\begin{document}

\title{Network reconstruction may not mean dynamics prediction}
\author{Zhendong Yu$^{1}$}
\author{Haiping Huang$^{1,2}$}
\email{huanghp7@mail.sysu.edu.cn}
\affiliation{$^{1}$PMI Lab, School of Physics,
Sun Yat-sen University, Guangzhou 510275, People's Republic of China}
\affiliation{$^{2}$Guangdong Provincial Key Laboratory of Magnetoelectric Physics and Devices,
Sun Yat-sen University, Guangzhou 510275, People's Republic of China}
\date{\today}

\begin{abstract}
	With an increasing amount of observations on the dynamics of many complex systems, it is necessary to reveal the underlying mechanisms behind these complex dynamics, which is fundamentally important in many scientific fields such as climate, financial, ecological, and neural systems. The underlying mechanisms are commonly encoded into network structures, e.g., capturing how constituents interact with each other to produce emergent behavior. Here, we address whether a good network reconstruction suggests a good dynamics prediction. The answer depends on the nature of the supplied (observed) dynamics sequences measured on the complex system. When the dynamics are not chaotic, network reconstruction implies dynamics prediction. In contrast, even if a network can be well reconstructed from the chaotic time series (chaos means that many unstable dynamics states coexist), the prediction of the future dynamics can become impossible as at some future point the prediction error will be amplified. This is explained using dynamical mean-field theory on a toy model of random recurrent neural networks.
\end{abstract}

 \maketitle

\section{Introduction}
Experimental technique advances provide us a huge amount of data across a broad range of scientific fields, such as cellular systems~\cite{Cell-2005}, ecological systems~\cite{Eco-2010}, economic systems~\cite{Ecophys-2014}, climate pattern dynamics~\cite{Climate-2021}, disease spreading dynamics~\cite{DS-2015}, and brain dynamics~\cite{Nature-2004,Nature-2022,Neuron-2024}. Dynamics is one salient feature of complex systems, and mastering the knowledge of dynamics allows one to predict the potential emergent behavior when external or internal conditions evolve~\cite{Pred-2012}. This plays a vital role in understanding neural correlates of cognition, tipping points of climate change, disease spreading, and financial crisis. There thus appear extensive research on network inference of hidden connections among constituents~\cite{Timme-2011,JCN-2015,Doya-2017,Rajan-2021,Grab-2023}, and even dynamics reconstruction~\cite{Ott-2018,Kim-2021,Zhao-2021,PRL-2023}. 

Most time series in nature are inherently chaotic~\cite{Chaos-2009}, especially those in a high dimensional state space, for example, an ongoing debate on chaotic fluctuations in spiking dynamics of neocortex~\cite{Ostojic-2014,F1k-2016}. By collecting spiking data from brain regions, one asks scientific questions about the cognitive behavior of brains~\cite{Nature-2022}, by finding the latent network structures supporting those observed dynamics~\cite{PRE-2016,IEEE-2016,Grab-2023}. This involves a fundamental question about whether the network dynamics can be forecasted once the network structure is well predicted. Previous studies highlighted the difficulty of predicting chaotic time series~\cite{Kanter-2002,Daniel-2022}, but a thorough theoretical understanding is lacking so far. 

Although in most complex systems, the evolution laws are unknown, we set up a simple model of recurrent dynamics ubiquitous in neural circuits. The synaptic couplings among neurons are pre-designed, and then the dynamics data are generated. Based on the toy dynamics data, we can implement the next-step prediction rule to reconstruct the network structure. An evaluation of the dynamics forecasting is carried out based on the inferred network. The dynamics phase can be tuned by a single synaptic gain parameter, and thus whether the network reconstruction implies the dynamics prediction can be addressed in this context. Importantly, we provide an analytic computation of the dynamics deviation between reconstructed and true networks by using the dynamical mean-field theory (DMFT).
\blue{We finally clarify when and how the prediction difficulty arises by defining a measure of the dynamics predictability given afford dynamics prediction error.}

\blue{We emphasize that in this work, we define and theoretically investigate the structural chaos (identical initial state but slightly different network structure), rather than the standard dynamics chaos with exactly the same network structure, by proposing a dynamical mean-field theory framework. This is in particular relevant in the current data-mining era of network inference and dynamics forecasting. We discuss in the last section the impacts on the current status of extracting latent dynamics equations from a massive amount of data in a variety of complex systems~\cite{Champ-2019,Brun-2015,Doya-2017,Rajan-2021}.  }

\section{Problem setting}\label{prob}
To generate a simulated time series, we first design a teacher recurrent neural network (RNN) composed of $N$ fully-connected neurons. 
The state of each neuron in time $t$ is characterized by the synaptic current $x_i^{(1)}(t)$ ($i= 1,\ldots,N$), which obeys the following first-order differential
 dynamics equation:
\begin{equation}\label{teacher_rnn}
 \frac{{dx_i^{(1)}}}{{dt}} =  - x_i^{(1)} + \sum\limits_{j = 1}^N {J_{ij}^{(1)}} \phi _j^{(1)}(t),
\end{equation}
where $\phi _j^{(1)}(t) = \tanh (x_j^{(1)}(t))$ transforms current to firing rate. \blue{We use $\{x_{i}\}$ to mean neual activity in the paper.} Each element $J_{ij}^{(1)}$ of the connection matrix is drawn from an independent Gaussian distribution as
\begin{equation}
   J_{ij}^{(1)} \sim \begin{cases}\mathcal{N}\left(0, \frac{g^2}{N}\right) & \text { for } i \neq j \\ 0 & \text { for } i=j\end{cases} .
\end{equation} 
Increasing the value of $g$ (the gain parameter) changes the dynamics phase from a global stable fixed point to a proliferation of an exponential number of unstable fixed points (deterministic chaos), while for a finite size, limit cycles of activity can also be observed around the transition point $g_c=1$~\cite{Huang-2022}. 

\begin{figure}
\centering
\includegraphics[width=0.85\textwidth]{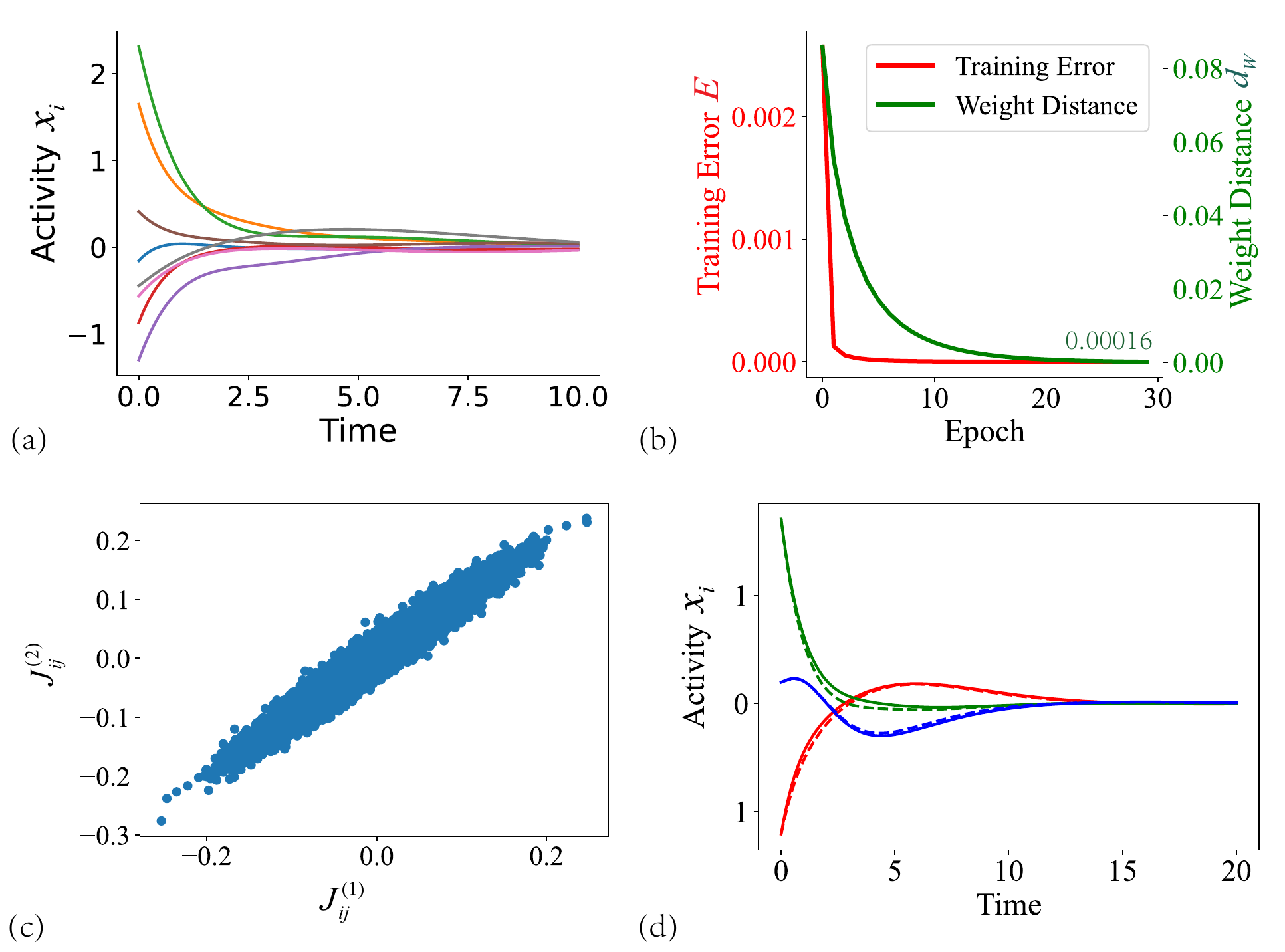}
\caption{Learning performance before the chaos transition with $N = 200$ and $g = 0.8$.
 (a) The activity trajectory of some selected neurons.
 (b) The training prediction (mean-squared) error \blue{$E(\{J_{ij}^{(2)}\})$ and the normalized Euclidean distance $d_{W}$} between the learned weights and the true weights.
 (c) Scatter plot of the learned weights (student) against the true weights (teacher). Only those with larger deviations are shown. (d) Student dynamics (dashed lines) match teacher dynamics (solid lines). Three typical neurons are shown.}
\label{fig1}
\end{figure}

\begin{figure}
\centering
\includegraphics[width=0.7\textwidth]{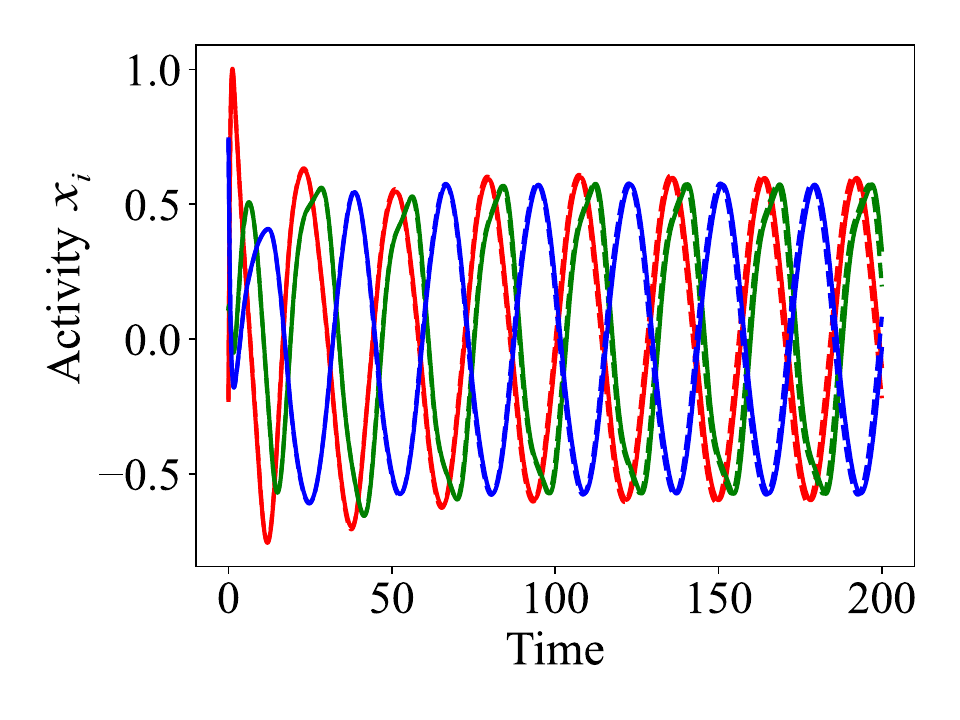}
\caption{ Learning performance in the limit cycle case $N = 200$ and $g = 1.2$.
 Student dynamics (dashed lines) match teacher dynamics (solid lines). Three typical neurons are shown.}
\label{fig2}
\end{figure}

 The dynamics in Eq.~\eqref{teacher_rnn} can be discretized into the following form:
\begin{equation}
x_i^{(1)}({t_l}) - x_i^{(1)}({t_{l - 1}}) =  - x_i^{(1)}({t_{l - 1}})h + \sum\limits_{j = 1}^N {J_{ij}^{(1)}\phi _j^{(1)}({t_{l - 1}})h}  + {\delta _{l,0}}{a_i},
\end{equation}
where ${\delta _{l,0}}$ represents the Kronecker symbol, $h$ is a small discretization step, and ${a_i} \overset{\text{i.i.d.}}\sim \mathcal{N}\left(0, 1\right) $ represents a Gaussian initial state unless otherwise specified.  By running the discrete-time dynamics, one can collect many trajectories starting from different initial points for the network reconstruction based on the next-step prediction rule. \blue{Using more advanced discretization methods such as the fourth-order Runge-Kutta method than this simple Euler scheme does not have a qualitative change of our following results.}

To learn the simulated temporal sequences, we design a student network with parameters $\mathbf{J}^{(2)}$, which obeys the same first-order differential equation:
\begin{equation}
\frac{d x_i^{(2)}}{d t}=-x_i^{(2)}+\sum_{j=1}^N J_{ij}^{(2)} \phi_j^{(2)}(t).\label{student_rnn}
\end{equation}
Suppose that the $J_{ij}^{(1)}$ is exactly recovered, we are able to predict the future dynamics of the teacher RNN. Now, we design the parameter learning as an optimization problem~\cite{IEEE-2016}, namely the next-step prediction with the following mean-squared error function:
\begin{equation}\label{NSP}
 E(\{J_{ij}^{(2)}\}) = \frac{1}{L}\sum\limits_{i,l} {{{\left[ {F_i[\bx^{(1)}({t_{l - 1}})] - x_i^{(1)}({t_l})} \right]}^2}} ,
\end{equation}
where $F_i[\bx^{(1)}({t_{l - 1}})] \equiv x_i^{(1)}({t_{l - 1}}) - x_i^{(1)}({t_{l - 1}})h + \sum\limits_{j = 1}^N {J_{ij}^{(2)}} \phi _j^{(1)}({t_{l - 1}})h$, and $L$ represents the total number of steps to be predicted in an example sequence. The first $500$ time steps of $100$ trajectories are collected as the training data, and the stochastic gradient descent using mini-batches and Adam optimizer are applied in the sequence learning~\cite{Adam}. \blue{The student can only have access to the time series generated by the teacher.}

\begin{figure}
\centering
\includegraphics[width=0.85\textwidth]{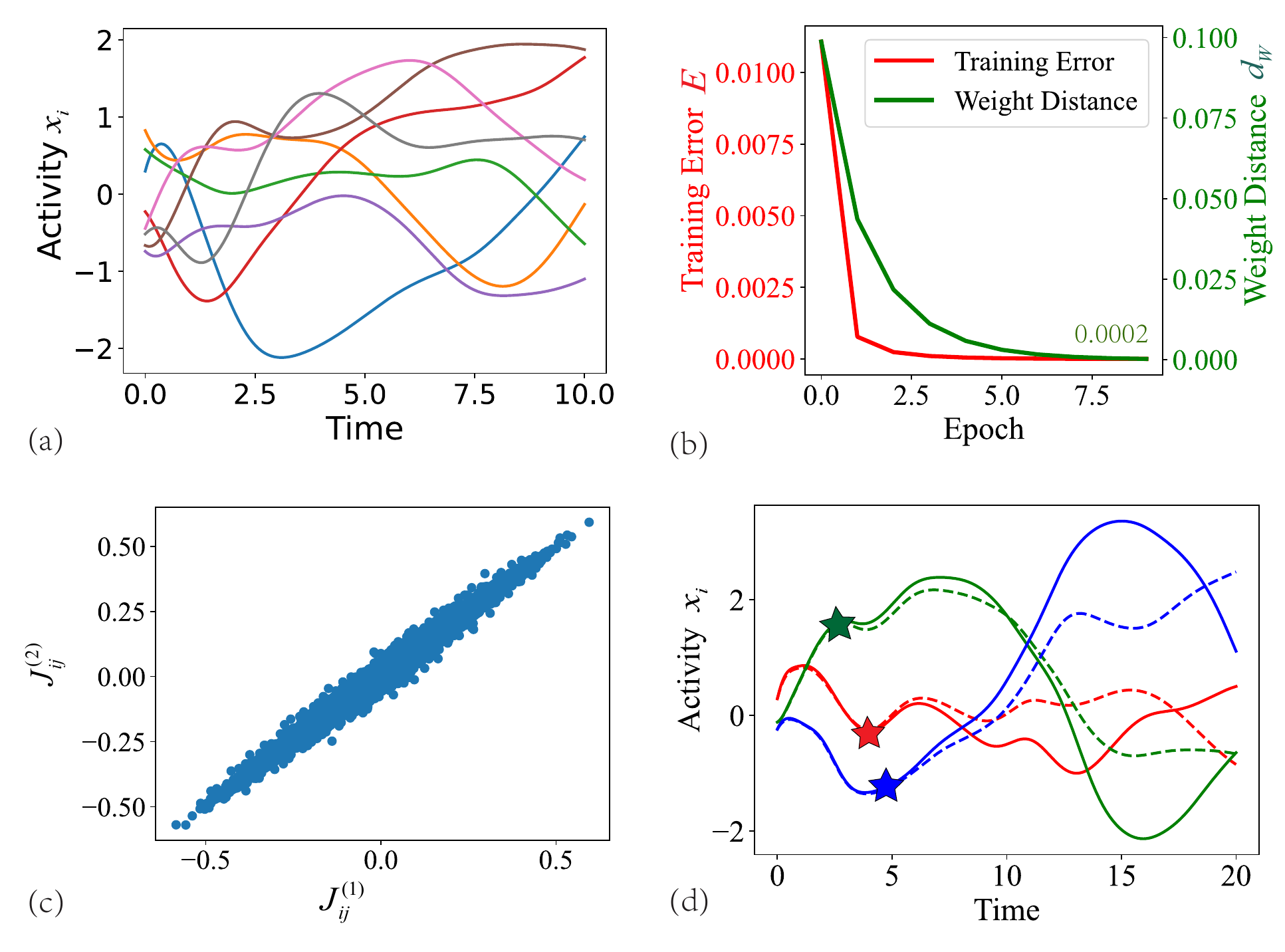}
\caption{ Learning performance after the chaos transition with $N = 200$ and $g = 2$.
 (a) The activity trajectory of some selected neurons.
 (b) The training prediction (mean-squared) error \blue{$E(\{J_{ij}^{(2)}\})$ and the normalized Euclidean distance $d_{W}$} between the learned weights and the true weights.
 (c) Scatter plot of the learned weights (student) against the true weights (teacher). Only those with larger deviations are shown.
 (d) Student dynamics (dashed lines) match teacher dynamics (solid lines) up to some time point (marked as a star). Three typical neurons are shown.}
\label{fig3}
\end{figure}

\section{Learning performance}\label{emp}
 We first show how the reconstruction of the teacher weight matrix behaves and whether the dynamics can be predicted after reconstruction. We consider two dynamics phases---null-activity and chaos phases. In the case of the trivial phase, both reconstruction and prediction are perfect (Fig.~\ref{fig1}), and for a finite-sized network, oscillatory dynamics may appear, and in this case, the reconstruction can still perfectly predict future dynamics of the limit cycles (see Fig.~\ref{fig2}), while in the chaotic phase, the network reconstruction by the next-step prediction still works well, i.e., both training error and weight distance can be reduced to a very low level (but not exactly zero), but the dynamics prediction based on the reconstructed weights works well only within an earlier stage of the dynamics, i.e., after some points [marked in Fig.~\ref{fig3} (d)], the student dynamics starts to deviate from the ground truth, and moreover, the deviation point shows a strong heterogeneity across neurons and are unpredictable.  Experimental details are given in the Appendix~\ref{app-a}.

\section{Dynamical mean field theory}
Dynamical mean field theory is a powerful approach for reducing high dimensional dynamics to low-dimensional (usually one or two) mean-field dynamics, for which a mechanistic interpretation is possible. The method was originally developed to understand spin dynamics in physics models~\cite{MSR-1973,Jans-1976,Domin-1978}. A pedagogical introduction of this method in the context of neural network dynamics was presented in a recent work~\cite{Zou-2024}.  

We consider the teacher model as the ground-truth model, whose dynamics are described by Eq.~\eqref{teacher_rnn}. To explain the observed unpredictability of the dynamics, we perturb the ground-truth $\mathbf{J}^{(1)}$ in the following way:
 \begin{equation}\label{decomp}
    J_{ij}^{(2)} = \sqrt {1 - {\eta ^2}} J_{ij}^{(1)} + \eta \delta J_{ij}^{(1)},
\end{equation}
which is the same as that used in a previous work studying the functional sensitivity of a random layered network~\cite{Li-2020}, $\eta$ indicates the perturbation strength, and 
$\delta J_{ij}^{(1)}$ is independent of $J_{ij}^{(1)}$ and satisfies the identical standard normal distribution:
\begin{equation}
   \delta J_{ij}^{(1)} \overset{\text{i.i.d.}}\sim \begin{cases}\mathcal{N}\left(0, \frac{g^2}{N}\right) & \text { for } i \neq j \\ 0 & \text { for } i=j\end{cases}.
\end{equation}
It can be proved that both $J_{ij}^{(2)}$ and $J_{ij}^{(1)}$ follow the same Gaussian distribution, and furthermore the angle $\theta_J$ between two weight vectors $\mathbf{J}^{(1)}$ and $\mathbf{J}^{(2)}$ is determined by $\sin(\theta_J)=\eta$.

Starting from \textit{an identical initial state} as in the teacher dynamics, the student dynamics obeys Eq.~\eqref{student_rnn}. The equivalent normalized Euclidean distance between two sets of weights ($\mathbf{J}^{(1)}$ and $\mathbf{J}^{(2)}$) in a finite-sized network is given by $d_W=2(1-\sqrt{1-\eta^2})g^2/N$. As long as $\eta\sim 0$, the deviation $d_W\sim\eta^2$.
Next, we ask how the normalized Euclidean distance between teacher and student dynamics evolves. The dynamics deviation can be decomposed into correlation functions amenable in the dynamical mean-field theory. This proceeds as follows:
\begin{equation}
\begin{aligned}
\frac{1}{N}\left\|\bx^{(1)}(t)-\bx^{(2)}(t)\right\|^2_2 & =\frac{1}{N} \sum_{i=1}^N\left[x_i^{(1)}(t)-x_i^{(2)}(t)\right]^2 \\
& =\frac{1}{N} \sum_{i=1}^N\left[x_i^{(1)}(t)\right]^2+\frac{1}{N} \sum_{i=1}^N\left[x_i^{(2)}(t)\right]^2-\frac{2}{N} \sum_{i=1}^N x_i^{(1)}(t) x_i^{(2)}(t) \\
& \simeq c^{11}(t, t)+c^{22}(t, t)-2 c^{12}(t, t),
\end{aligned}
\end{equation}
where we define the following correlation functions:
\begin{equation}
    \frac{1}{N} \sum_{i=1}^N x_i^{({\alpha})}(t) x_i^{({\beta})}(s) \simeq c^{{\alpha} {\beta}}(t, s).
\end{equation}
The above approximation becomes exact in the thermodynamic limit. We remark that ${c^{11}}(t,t) = {c^{22}}(t,t)$ at any time, since in statistics, the student and teacher bear the same coupling distribution.

We then write the deterministic dynamics into a path integral form (a generating functional of the correlation functions in physics) and then average out the disorder in the coupling statistics. Finally, we derive the following dynamical mean-field equation \blue{(i.e., the DMFT solution of the dynamics forecasting problem)}:
\begin{equation}\label{dmft}
    \left(\partial_t+1\right) x^{(\alpha)}(t)=\gamma^{(\alpha)}(t) \quad {\alpha} \in\{1,2\},
\end{equation}
where
\begin{subequations}
\begin{align}
   &\left\langle {{\gamma ^{(\alpha)} }(s){\gamma ^{(\beta)} }(t)} \right\rangle  = {g^2}\left\langle {{\phi ^{(\alpha)} }(s){\phi ^{(\beta)} }(t)} \right\rangle [(1 - {\delta _{{\alpha} {\beta} }})\sqrt {1 - {\eta ^2}}  + {\delta _{{\alpha} {\beta} }}], \\
   &    {\phi ^{(\alpha)} }(s) = \tanh [{x^{(\alpha)} }(s)],\\
   &{x^{(1)}}(0) = {x^{(2)}}(0) = a,\\
   &a \sim \mathcal{N}(0,1).
\end{align}
\end{subequations}
The detailed derivation of the above dynamical mean-field dynamics is given in Appendix~\ref{app-b}. The two-dimensional mean-field dynamics can be numerically solved~\cite{Zou-2024}, and numerical details are given in Appendix~\ref{app-c}.

As derived before, the deviation of the student dynamics from the teacher one can be tracked by estimating the following function:
\begin{equation}
\begin{aligned}
    d(t) &= \left\langle {{x^{(1)}}(t){x^{(1)}}(t)} \right\rangle  + \left\langle {{x^{(2)}}(t){x^{(2)}}(t)} \right\rangle  - 2\left\langle {{x^{(1)}}(t){x^{(2)}}(t)} \right\rangle,\\
    &=2[c^{11}(t,t)-c^{12}(t,t)].
    \end{aligned}
\end{equation}
Note that $c^{\alpha\beta}(t,t)$ can be obtained by solving the two-dimensional coupled mean-field dynamics equation (see Eq.~\eqref{dmft}, and Appendix~\ref{app-c}). Therefore, the empirical behavior in section~\ref{emp} can be theoretically explained.

By analogy with the Lyapunov exponent in characterizing the dynamics stability~\cite{Huang-2022,Jiang-2023}, we can also generalize this concept to our dynamics prediction case.
In this context, despite the same initial condition for the dynamics, we perturb the teacher's weights, corresponding to the fact that a latent network can not be reconstructed in a manner of exactly zero error, which is attributed to the heuristic algorithms used to predict the latent weights on one hand, and the numerical accuracy on the other hand. Hence, the potential dynamics deviation is caused by a slight difference between the ground truth and the approximately inferred one. This weight difference can be calculated in our case $d_W=\frac{g^2\eta^2}{N}$ in the small $\eta$ limit. We can thus define the Lyapunov exponent for the dynamics prediction as follows,
\begin{equation}\label{lydp}
    \lambda_{\text {dp }}=\lim _{t \rightarrow \infty,\eta\to 0} \frac{1}{t} \ln \left(\frac{\|\delta \mathbf{x}(t)\|_2}{\epsilon^{1/2}(\|d \J\|^2_2)}\right),
\end{equation}
where $\|\delta \mathbf{x}(t)\|^2_2=\sum\limits_i {{{(x_i^{(1)}(t) - x_i^{(2)}(t))}^2}}=Nd(t)$, and \blue{$\|d\J\|^2_2 = \sum\limits_{i,j} {{{(J_{ij}^{(1)} - J_{ij}^{(2)})}^2}}\simeq N^2d_W$} which induces the first-step deviation $\epsilon$ of dynamics \blue{($\epsilon=\|\bx^{(1)}(t=1)-\bx^{(2)}(t=1)\|_2^2$, where $t=1$ means the first-step dynamics of the student and teacher networks given the same initial condition; see more technical details in Appendix~\ref{app-d})}. Note that the initial condition is exactly the same for the two trajectories, such that $\epsilon=0$ if $\eta=0$, but a small value if $\eta$ is an infinitesimal value. Therefore, the possible deviation is caused by a cooperation of both $\eta$ and the first-step deviation. \blue{Equation~\eqref{lydp} thus defines a generalized Lyapunov exponent for possible \textit{structural} chaos in the context of network-inference and dynamics-prediction.}

 The dynamics-prediction Lyapunov exponent also yields in turn a measure of the dynamics predictability $T_{\rm pd}\equiv\frac{1}{\lambda_{\rm dp}}\ln\left(\frac{\delta}{\epsilon^{1/2}}\right)$ (see a similar definition in a related context~\cite{Pred-2012}), where $\delta$ means the afford
prediction error.  $T_{\rm pd}$ tells us the prediction time window is limited, or a long-term prediction in the presence of slight model deviation is impossible when the model is learned from high-dimensional chaotic time series.

\begin{figure}
\centering
\includegraphics[width=0.85\textwidth]{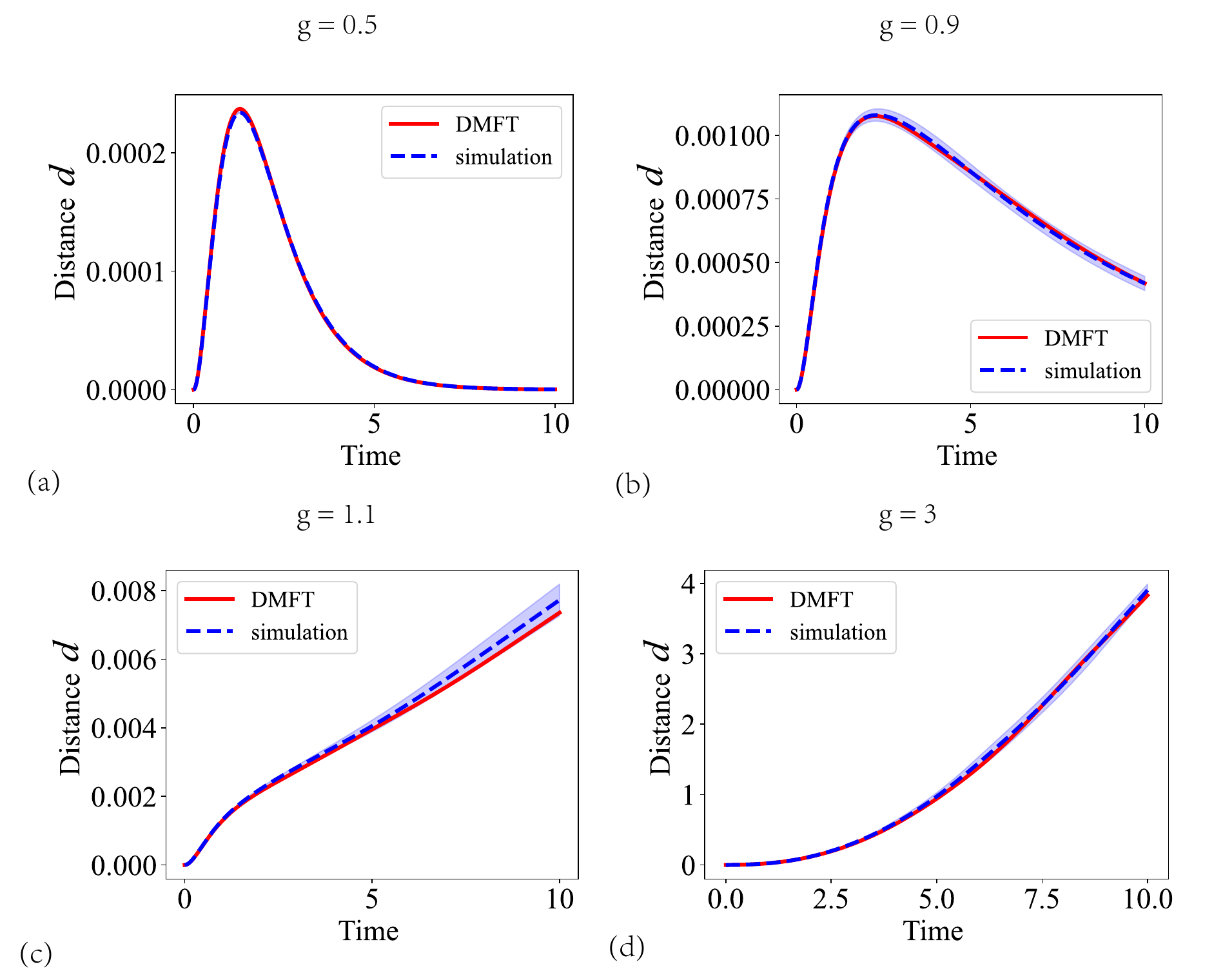}
\caption{Dynamical mean-field results \blue{[the trajectory distance $d(t)$]} for the perturbation $\eta = 0.1$.  The results are obtained by solving the DMFT equations for different values of $ g$:  (a) $g = 0.5$; (b) $g = 0.9$; (c) $g = 1.1$; and (d) $g = 3$. We then compared these results with simulations of RNN dynamics in a system with $10\,000$ neurons, where an average over five independent trials is done.}
\label{fig4}
\end{figure}

\begin{figure}
\centering
\includegraphics[width=0.85\textwidth]{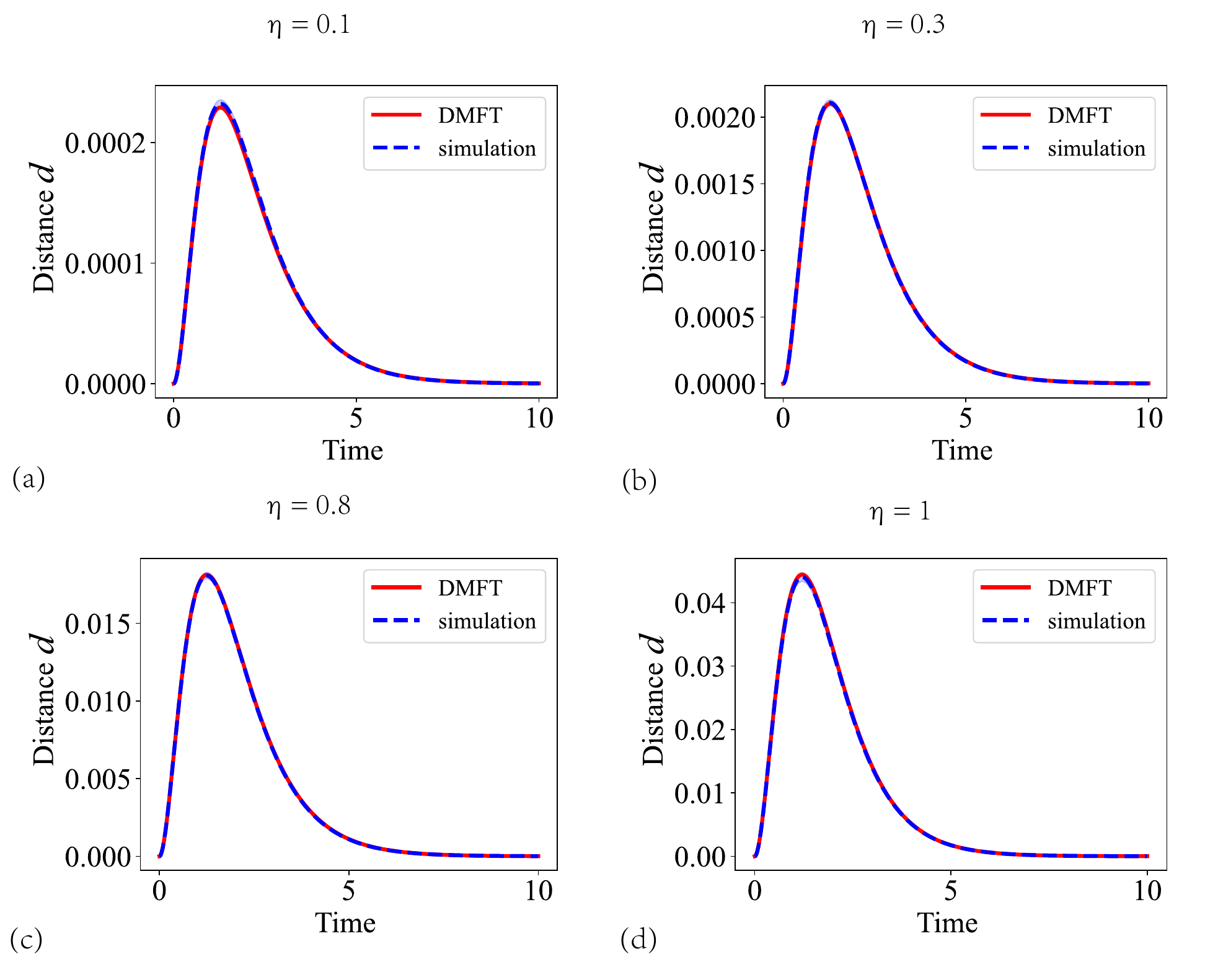}
\caption{Dynamical mean-field results \blue{[the trajectory distance $d(t)$]} for $g=0.5$ yet with different perturbation strength $\eta$. (a) $\eta=0.1$; (b) $\eta=0.3$; (c) $\eta=0.8$; (d) $\eta=1$.
 We then compared these results with simulations of RNN dynamics in a system with $10\,000$ neurons, where an average over five independent trials is done.}
\label{fig5}
\end{figure}

\section{Theoretical results}
\begin{figure}
\centering
\includegraphics[width=0.85\textwidth]{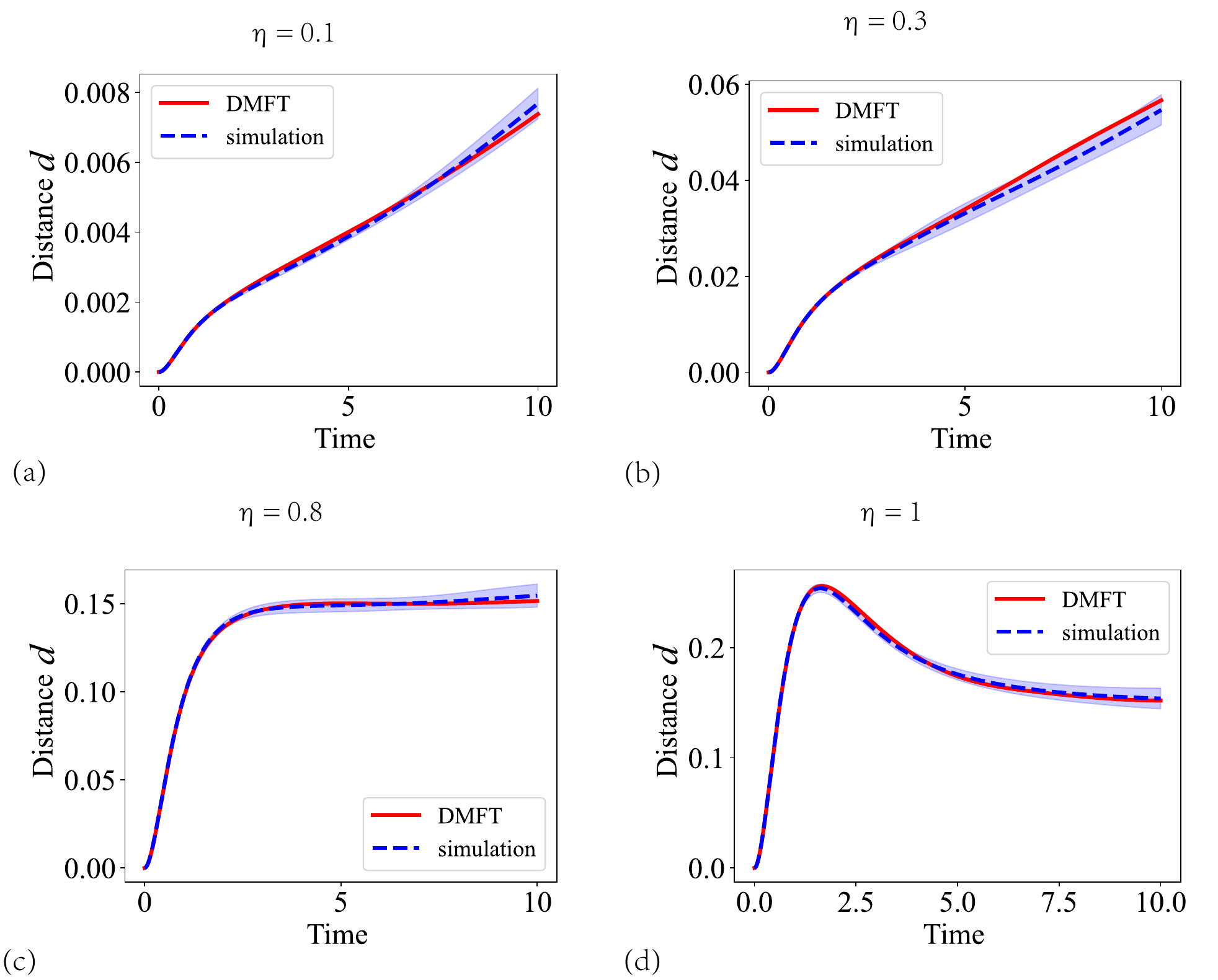}
\caption{Dynamical mean-field results [the trajectory distance $d(t)$] for $g=1.1$ yet with different perturbation strength $\eta$.
(a) $\eta=0.1$; (b) $\eta=0.3$; (c) $\eta=0.8$; (d) $\eta=1$.
 We then compared these results with simulations of RNN dynamics in a system with $10\,000$ neurons, where an average over five independent trials is done.}
\label{fig6}
\end{figure}

Theoretical predictions are plotted against the numerical simulations of finite-sized networks ($N=10\,000$ neurons) in Fig.~\ref{fig4}.
As the gain parameter increases, the profile of the trajectory distance has a qualitative change at the transition point $g_c=1$. Before the transition, the distance converges to zero, demonstrating that $\lambda_{\rm pd}$ is not positive. After the transition, the distance behavior changes qualitatively to a growing fashion with time, suggesting a positive $\lambda_{\rm pd}$. We thus conclude that in this chaotic regime, the dynamics can not be well predicted unless the coupling can be perfectly recovered, which is quite difficult due to the existence of numerical noise (an equivalent small amount of $\eta$ in our model). In Fig.~\ref{fig5} and Fig.~\ref{fig6}, we show the comparison between theoretical prediction and simulation results in trivial and chaos phases, respectively.

\begin{figure}
\centering
\includegraphics[width=0.85\textwidth]{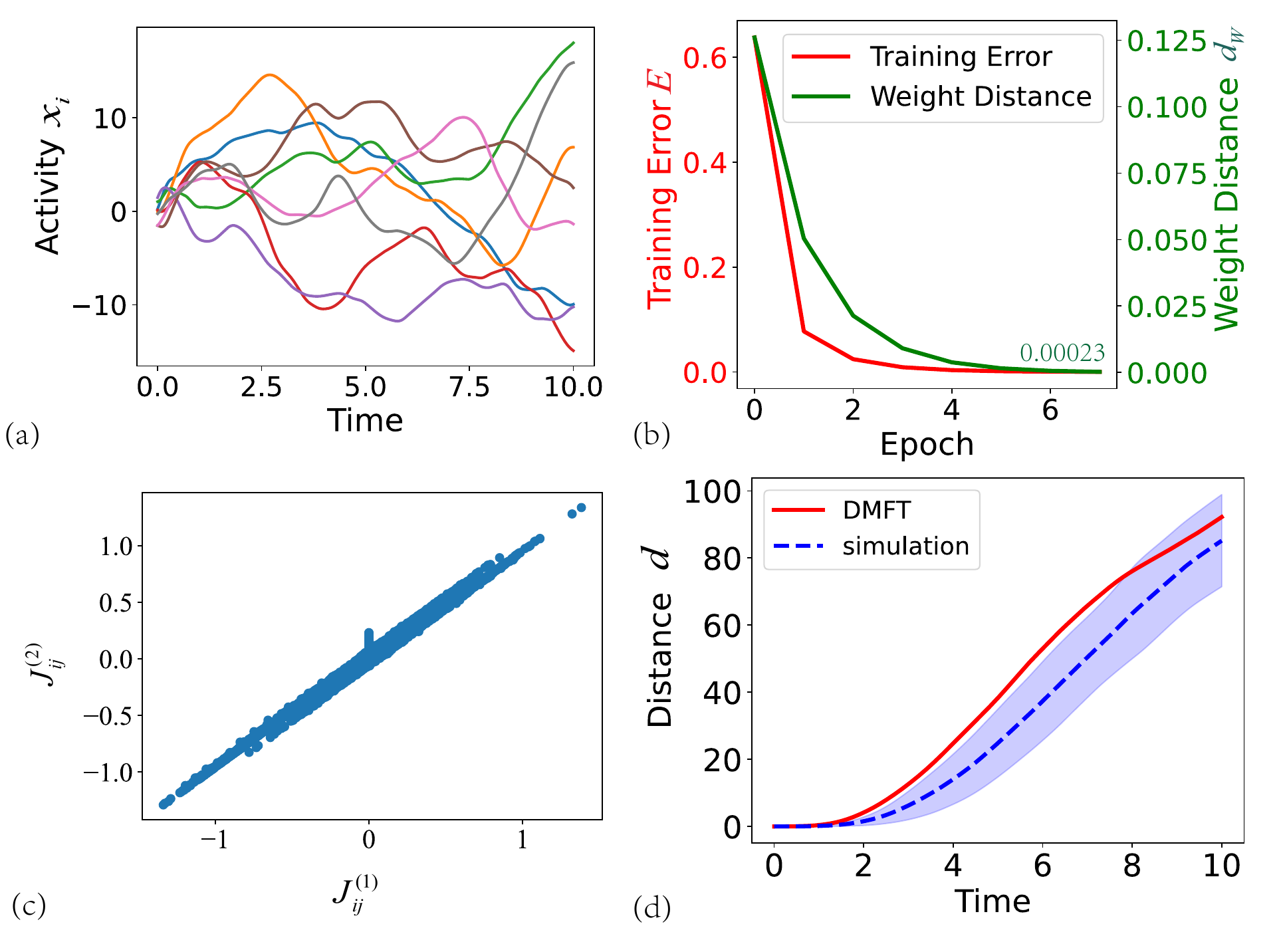}
\caption{Theory compared with practical network reconstruction. $N = 1\,000$ , $g = 10$, and $\eta  = 0.048$.
 (a) Typical trajectories of neural currents.
 (b) The training error \blue{$E(\{J_{ij}^{(2)}\})$ and the normalized Euclidean distance $d_{W}$} between learned and true weights.
 (c) Scatter plot of student weights against teacher ones. Only those with larger deviations are shown. (d) Dynamics deviation \blue{$d(t)$} predicted by the theory and compared with simulations (averaged over
 five independent trials).}
\label{fig7}
\end{figure}

\begin{figure}
\centering
\includegraphics[width=0.85\textwidth]{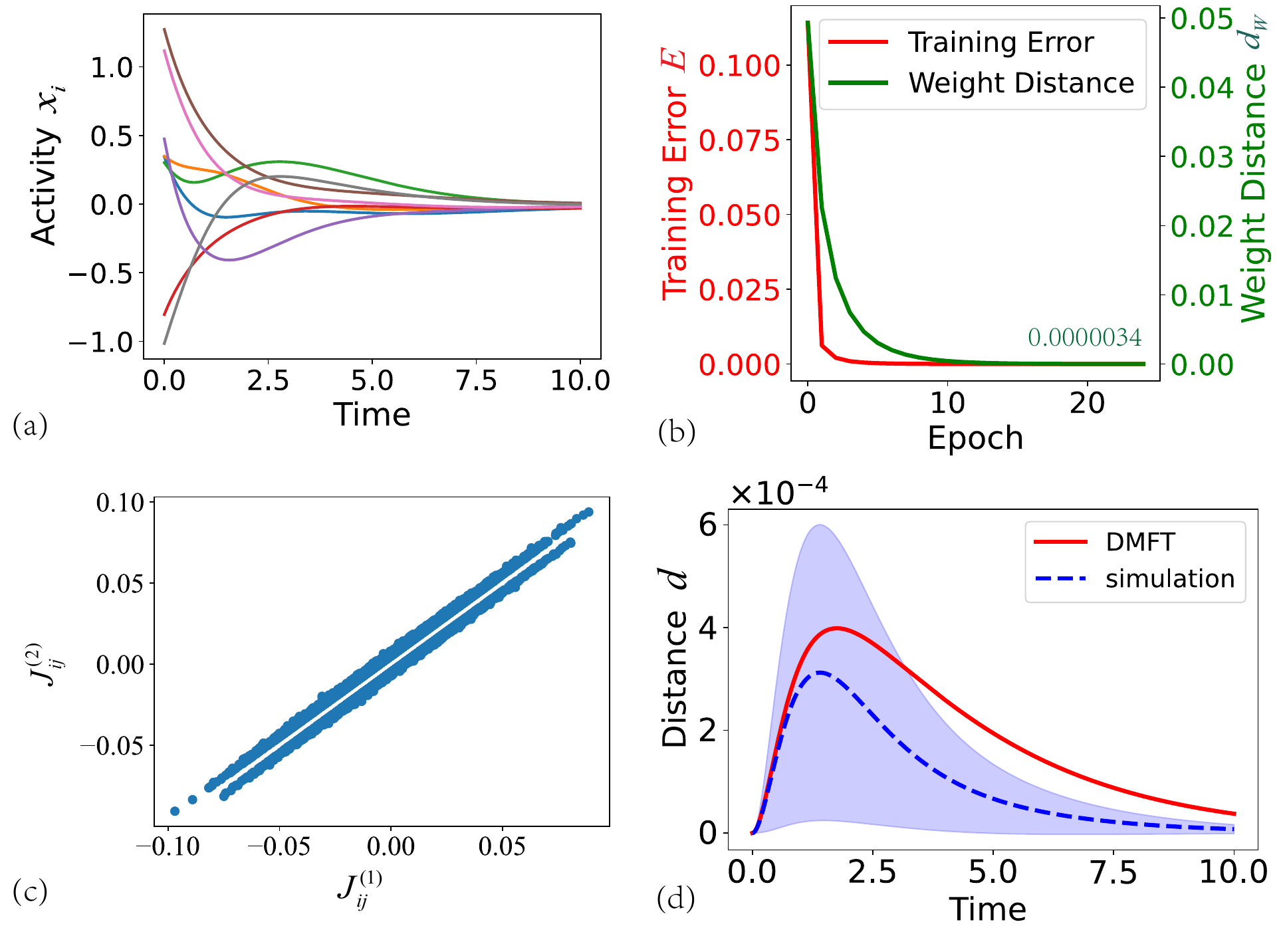}
\caption{
 Theory compared with practical network reconstruction. $N = 1\,000$ , $g = 0.8$, and $\eta  = 0.073$.
 (a) Typical trajectories of neural currents.
 (b) The training error \blue{$E(\{J_{ij}^{(2)}\})$ and the normalized Euclidean distance $d_{W}$} between learned and true weights.
 (c) Scatter plot of student weights against teacher ones. Only those with larger deviations are shown. (d) Dynamics deviation \blue{$d(t)$} predicted by the theory and compared with simulations (averaged over
 five independent trials). }
\label{fig8}
\end{figure}

Next, we establish the correspondence between the practical network reconstruction using the algorithm in Sec.~\ref{prob} and the dynamical mean-field theory. The key is to transform the perturbation $\eta$ back to the reconstruction weight distance, which is carried out as follows,
\begin{equation}\label{etafun}
\eta=\sqrt{1-\left(1-\frac{Nd_W}{2g^2}\right)^2},
\end{equation}
where $d_W$ is the normalized weight distance. \blue{The above relationship between $\eta$ and $d_{W}$ requires a small weight deviation such that the decomposition in Eq.~\eqref{decomp} can be used to derive Eq.~\eqref{etafun}.} As shown in Fig.~\ref{fig7} and Fig.~\ref{fig8}, our DMFT results predict qualitatively well the experimental behavior.

Next, we numerically estimate $\lambda_{\rm pd}$ as a function of the synaptic gain $g$. Technical details are given in Appendix~\ref{app-d}. Figure~\ref{fig9} shows that as $g$ crosses the critical value, the dynamics-prediction Lyapunov exponent becomes positive, demonstrating that the ability to predict future dynamics is immediately lost. Therefore, we reveal the intrinsic difficulty of predicting future chaotic behavior based on network reconstruction. A qualitative identification of dynamic collective behavior is relatively easy, as long as the inference error of model parameters is small, while a quantitative forecasting of a specific trajectory would become impossible if the trajectory is high-dimensional and chaotic (see the classic Lyapunov exponent in the upper inset of Fig.~\ref{fig9}), especially in a long-term prediction sense. This is explained by our DMFT equations in the high-dimensional limit. As the number of degrees of freedom is lowered down, the prediction may be improved in quality using advanced model-free methods or specially designed methods~\cite{Ott-2018,PRL-2023}. However, in a general context, the prediction from the past trajectories is inherently challenging.

\begin{figure}
\centering
\includegraphics[width=0.85\textwidth]{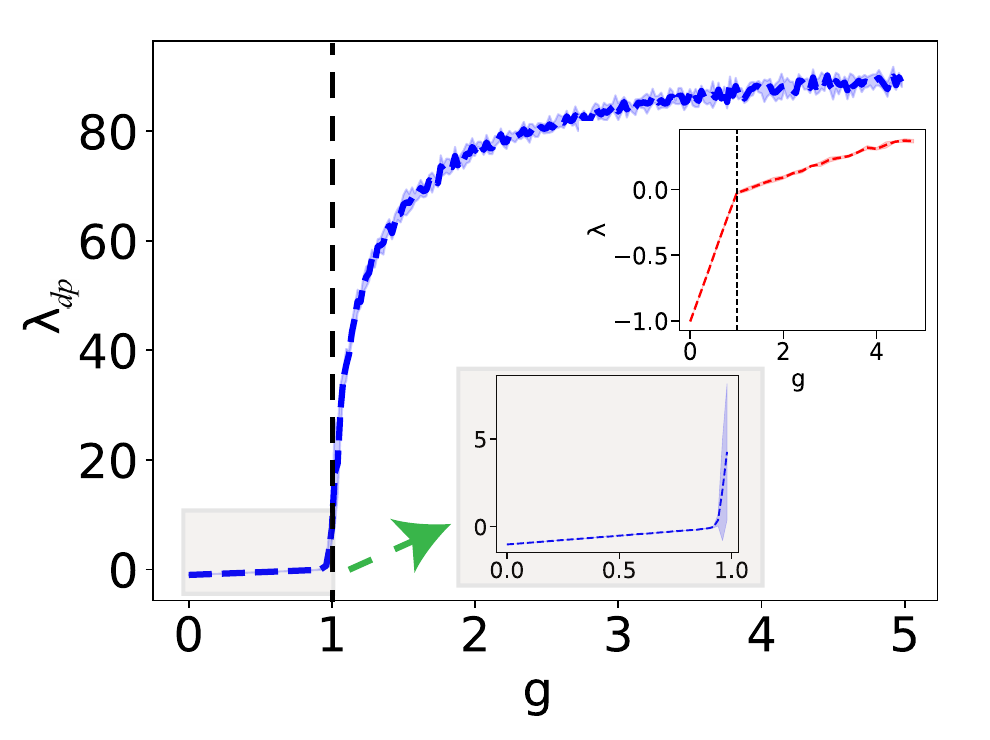}
\caption{ Lyapunov exponent for dynamic prediction as a function of synaptic gain $g$. $N = 1000$, $\eta  = 0.01$, and the discretization step size $h = 0.01$. The right upper inset shows the classic Lyapunov exponent, while the right lower one is an enlarged plot of the part $g<1$ in the main plot. The results are obtained from simulations on five individual trials. The results are qualitatively robust when the hyper parameters are chosen in the physically reasonable regime.  }
\label{fig9}
\end{figure}

 \begin{figure}
    \centering
    \subfigure[]{
        \includegraphics[width=0.48\textwidth]{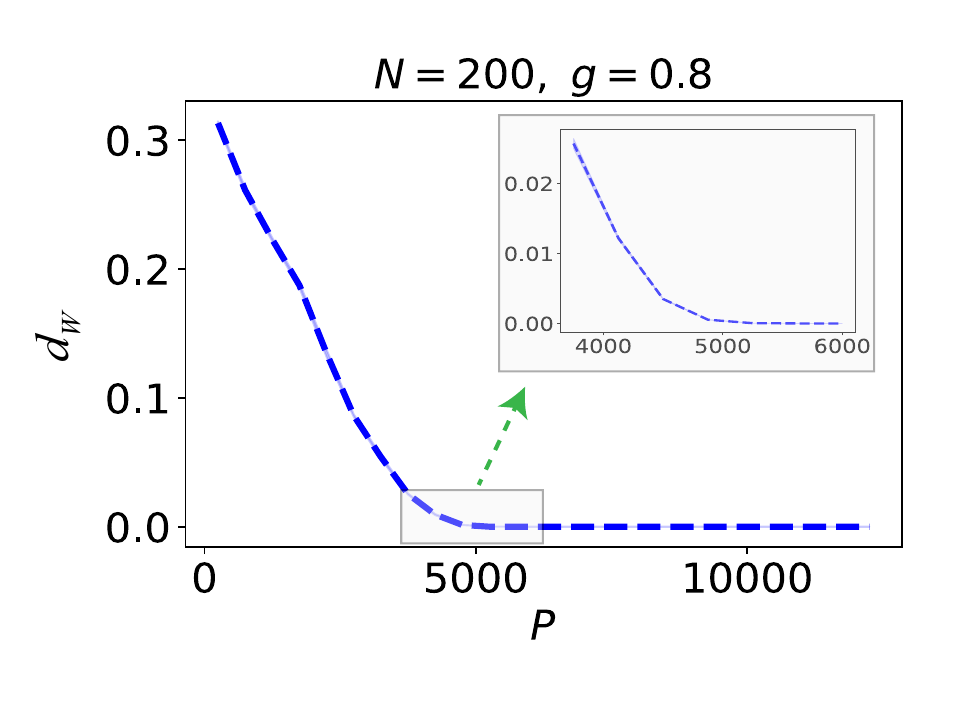}
   }
   \subfigure[]{
        \includegraphics[width=0.48\textwidth]{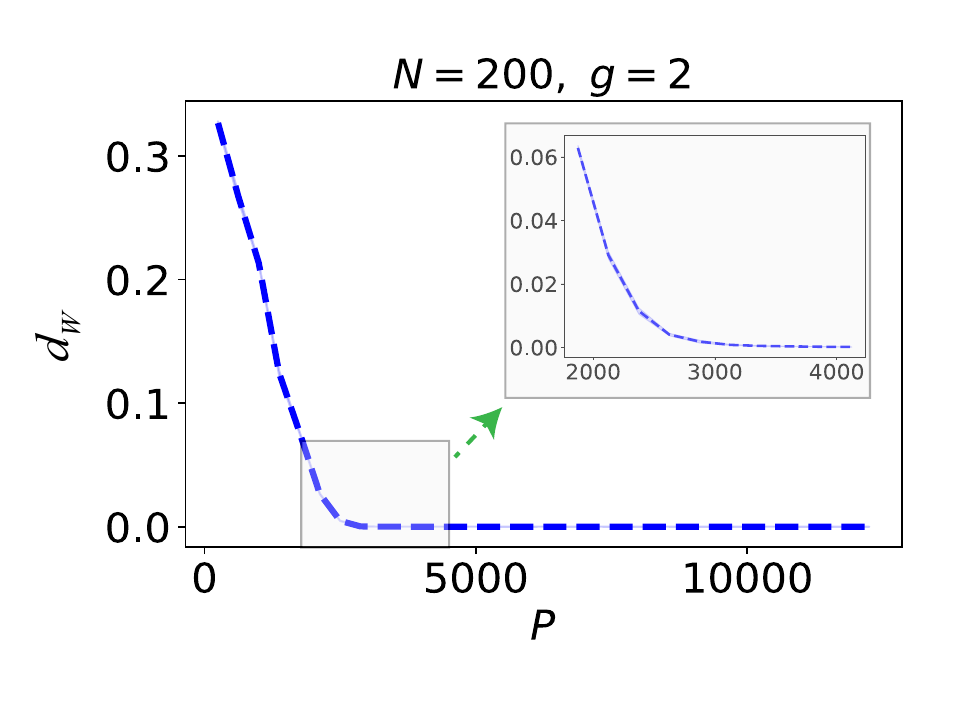}
       }
    \caption{\blue{The network strucuture deviation as a function of the number of training segments ($P=H\times S$, see Appendix.~\ref{app-a}). (a) non-chaotic data. (b)
    chaotic data. Five independent trials are considered to draw the plot.} }\label{fig10}
\end{figure}

\begin{figure}
\centering
\includegraphics[width=0.8\textwidth]{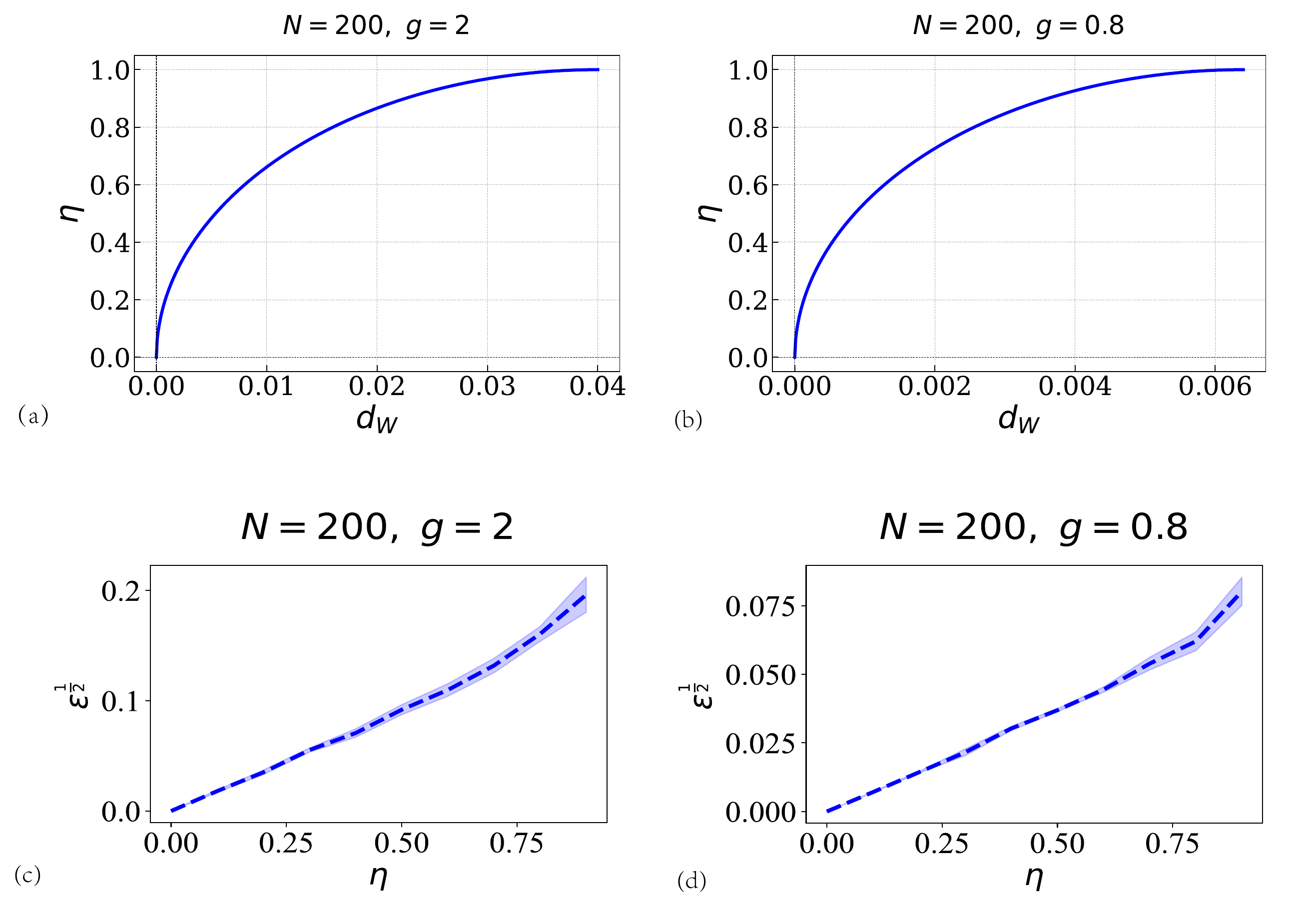}
\caption{\blue{The relationship among the first-step deviation of the dynamics $\epsilon$, the rotation magnitude $\eta$ and the inferred network deviation $d_{W}$. 
(a,b) $\eta$ as an analytic function of $d_{W}$ (see Eq.~\eqref{etafun} in the main text). (c,d) the first-step deviation as a function of $\eta$. This relationship can only be numerically evaluated (see technical details in Appendix~\ref{app-d}).}
}
\label{fig11}
\end{figure}

\begin{figure}
\centering
\includegraphics[width=0.85\textwidth]{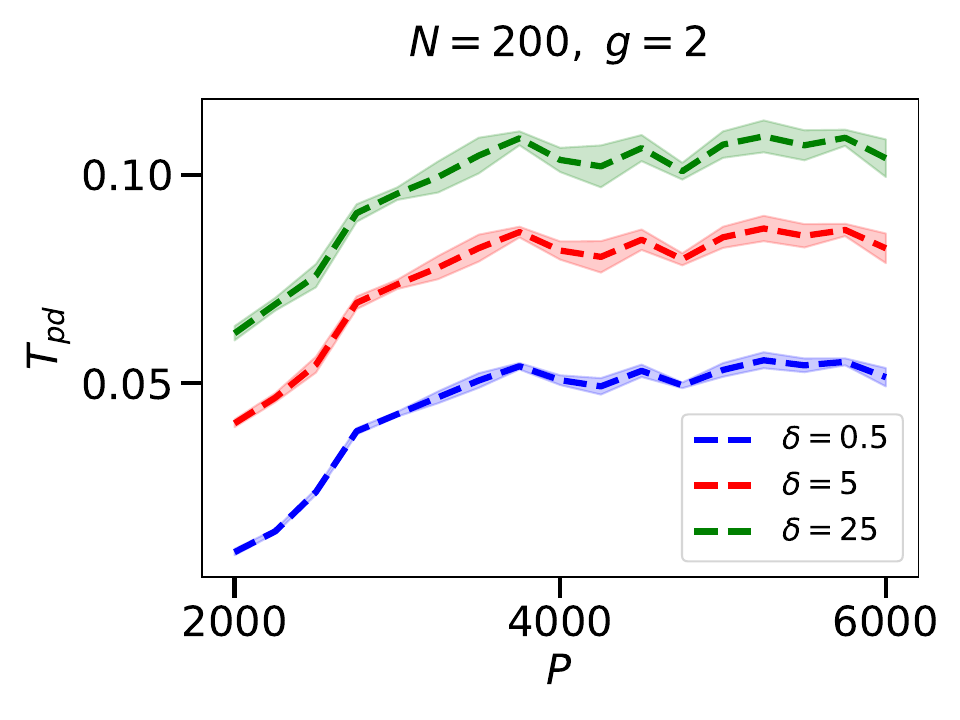}
\caption{\blue{The relationship between the dynamics predictability and the training data size $ P $. Three afford dynamics prediction errors are considered: $ \delta = 0.5$, $5$, and $ 25$. 
Five independent trials are considered.}}
\label{fig12}
\end{figure}

\blue{Finally, we show how the number of training examples affects the dynamics predictability specified by $T_{\rm pd}$. As shown in Fig.~\ref{fig10}, the deviation of the
student weight from the teacher weight shrinks with increasing training segments. According to Eq.~\eqref{etafun}, we can transform the network deviation $d_{W}$ to the rotation angle determined by $\eta$ [see Fig.~\ref{fig11} (a,b)]. After that, the rotation magnitude $\eta$ affects the first-step deviation $\epsilon$ between the dynamics of the student and the teacher [see Fig.~\ref{fig11} (c,d)]. Given an afford dynamics-prediction error $\delta$, the typical value of the prediction time window can be estimated by $T_{{\rm dp}}=\frac{1}{\lambda_{\rm dp}}\ln\left(\frac{\delta}{\epsilon^{1/2}}\right)$, which is shown in Fig.~\ref{fig12}. Note that given $N$, $g$ and $\eta$, the Lyapunov exponent for the dynamics prediction $\lambda_{\rm dp}$ can be evaluated using the method detailed in Appendix~\ref{app-d}. The dynamics predictability increases first with the training example yet gets saturated. As expected, lowering the prediction error decreases the prediction time window as well. The prediction time window is the time point up to which the dynamics prediction falls within the afford error (see Fig.~\ref{fig3} (d) as an example).
 }

\section{Conclusion}
In this paper, we address a fundamental question about whether a good reconstruction of latent network weights implies a good prediction of the dynamics. This question depends on the nature of the training dynamics sequence. We clarify the underlying physics by designing a toy model of network dynamics and then derive the dynamical mean-field theory explaining the experimental observations. More precisely, we generate an ensemble of teacher dynamics, which is used to learn the latent weight matrix of the teacher network. After the weight matrix is learned by a counterpart student network, the dynamics of the teacher starting from an arbitrary initial point are compared with the student prediction (the same initialization but different network topology). We find that, if the dynamics are posited in an order phase, such as fixed points or limit cycles, the network can be well reconstructed, and the dynamics can also be well predicted. In contrast, if the dynamics are chaotic, i.e., sensitive to a tiny random perturbation of the neural state, we can still reconstruct the network weight from these chaotic sequences, based on the principle of next-step prediction. However, due to the intrinsic (small) deviation from the ground-truth network weight, the dynamics prediction error can be amplified at some later moment of the trajectory. This phenomenon is predicted by our dynamical mean-field theory, where we equivalently introduce a rotation of the ground-truth weight and study how both dynamics deviate from each other given the identical initial state. Hence, we conclude that network reconstruction does not mean dynamics prediction, which depends on the nature of the supplied sequence data.

The conclusion we draw may shed light on current data-mining experiments across a broad range of scientific fields~\cite{Eco-2010,Ecophys-2014,Climate-2021,DS-2015,Doya-2017,Neuron-2024}, such as ecological dynamics, epidemiology, weather forecasting, and brain dynamics, etc. It is common to use many advanced techniques such as machine learning to extract latent structures underlying a huge amount of time sequences, e.g., those in weather time series, stock price dynamics, and cortical electrodynamics. Our study suggests that even if a latent ground-truth model exists, the future dynamics of the model (usually a natural complex system) could not be always predicted based on the inferred latent structure. At least in the example studied here, the dynamics are not predictable when the supplied data for the network reconstruction is chaotic, \textit{even if the network structure is inferred with a very high accuracy} (corresponding to a very small $\eta$ in our theoretical model). \blue{This is actually a structural chaos in dynamics forecasting (identical initial state but slightly different network structure), rather than the dynamics chaos in the standard definition through a tiny perturbation of the initial condition. We further demonstrate how the number of training examples impacts the dynamics predictability, providing a precise guideline for network reconstruction and dynamics forecasting in our problem setting. }

In fact, many dynamical systems may not have a well-documented evolution rule, and model-free machine learning methods are used to implement the prediction (always based on a black box of a gigantic amount of parameters, such as deep neural networks).  Our theoretical study thus asks further questions of how we can minimize the prediction error of the future dynamic behavior (even if this error could not be made zero) by designing more advanced techniques of statistical inference \blue{(e.g., using deep neural networks), and what limits the dynamics prediction in real-world applications (the nature of the time series or the algorithms, or both)}. Most importantly, it should be careful to draw a conclusion about the future behavior of dynamically adaptive complex systems, such as brain circuits~\cite{Nat-2017,Doya-2017,Rajan-2021}, even if a nearly-perfect construction of the circuit connectome is possible~\cite{Con-2023}. \blue{As recurrent neural network models are commonly used as a data mining tool for brain circuits~\cite{Nat-2017,Rajan-2021}, our theoretical work of the structural chaos would shed light on the relationship between network inference and dynamics forecasting in the field of network neuroscience.}

\section*{Acknowledgments}
We thank Zhanghan Lin for discussions at earlier stages of this project. This research was supported by the National Natural Science Foundation of China for
Grant number 12475045 and 12122515, and Guangdong Provincial Key Laboratory of Magnetoelectric Physics and Devices (No. 2022B1212010008), and Guangdong Basic and Applied Basic Research Foundation (Grant No. 2023B1515040023).

\onecolumngrid
\appendix
\section{Experimental details of the network structure inference}\label{app-a}

To prepare the training sequences, we first simulate the following discretized equation:
\begin{equation}
x_i^{(1)}({t_l}) - x_i^{(1)}({t_{l - 1}}) =  - x_i^{(1)}({t_{l - 1}})h + \sum\limits_{j = 1}^N {J_{ij}^{(1)}\phi _j^{(1)}({t_{l - 1}})h}  + {\delta _{l,0}}{a_i}.
\end{equation}
Firstly, an initial point $\mathbf{a}=\{a_i\}$ is sampled from an $N$-dimensional standard Gaussian distribution (a diagonal covariance matrix). We then set the time increment $ h = 0.01 $, and run the equation starting from the initial point for a total of $1\,000$ time steps to generate a trajectory. The first $500$ steps of this trajectory are then taken as one sample or training sequence. This process is repeated $100$ times to obtain $100$ samples. Each sample is subsequently divided into $125$ equal-interval segments, resulting in a total of $12\,500$ training segments. These segments are used to minimize the next-step prediction loss function [Eq.~\eqref{NSP}], with a mini-batch size of $250$. The learning rate is set to $0.1$. During training, the Adam optimizer is utilized. These hyper parameters including the experiment setup are used for simulating dynamics of RNN with $N=200$. For other network sizes, the procedure is similar. Codes are available in our GitHub~\cite{Yu-2024}.
\blue{The pseudocode is given in Alg.~\ref{alg1}. Hyperparameters used in figures are specified in Table~\ref{tab}. }

\begin{algorithm}
\caption{Trajectory generation and training procedure for RNN reconstruction}\label{alg1}
\begin{algorithmic}[1]
\State \textbf{Hyperparameters:} gain parameter $g$;
network size $N$, time increment $h$; total time steps $T_0$; truncated time steps $T$;
 the number of sampled trajectories $S$; learning rate ${\rm lr}$; mini-batch size $B$; 
 the number of equal-interval segments $H$; the number of epochs $M$.

\For{$i = 1$ to $S$} \Comment{Repeat $S$ times to generate $S$ samples}
    \State Sample initial point $\mathbf{a}_0 \sim \mathcal{N}(0, \id_N)$
    \State Run the RNN dynamics for $T_0$ steps with a time step size of $h$
    \State Trucate the first $T$ time steps for training and the remaining data for prediction
    \State Cut the recorded trajectory into $H$ equal-interval segments
    \State Store the segments for training
    \EndFor
\State \textbf{Input:} Stored training data
\State Initialize the Adam optimizer
\State \textbf{Training:} Set the learning rate to ${\rm lr}$ and the mini-batch size to $B$;
 Minimize the next-step prediction loss [Eq.~\eqref{NSP}] using stochastic gradient descent by running $M$ epochs.
\State \textbf{Output:} Reconstructed network 
\end{algorithmic}
\end{algorithm}

\begin{table}[H]
\centering
\caption{Hyperparameters for figures in the main text}\label{tab}
\begin{ruledtabular}
\begin{tabular}{ccccccccccc}
Figure & $g$   & $N$   & $h$   & $T_0$ & $T$   & $S$   & $\text{lr}$   & $B$   & $H$   & $M$   \\
\hline
1      & 0.8   & 200   & 0.01  & 1000  & 500   & 100   & 0.1   & 250   & 125   & 30    \\
3      & 2.0   & 200   & 0.01  & 1000  & 500   & 100   & 0.1   & 250   & 125   & 10    \\
7      & 10.0  & 1000  & 0.01  & 1000  & 500   & 300   & 0.02  & 250   & 125   & 8     \\
8      & 0.8   & 1000  & 0.01  & 1000  & 300   & 800   & 0.02  & 250   & 100   & 25    \\
10 (a)      & 0.8   & 200  & 0.01  & 1000  & 500   & -  & 0.1  & 250   & 125   & 10000    \\
10 (b)      & 2.0   & 200  & 0.01  & 1000  & 500   &  - & 0.1  & 250   & 125   &10000    \\
\end{tabular}
\end{ruledtabular}
\end{table}

\section{Derivation of dynamical mean-field equations}\label{app-b}
To compute analytically the deviation $d(t)$, we must first get the correlation functions $c^{\alpha\beta}(t,t)$, which can be in principle computed from a derivative of the path integral action function of the dynamics. In this section, we detail these steps.

We first set the discrete time steps $t_l=lh$, $l=0,1,\ldots,M$, such that $t_M=T$ (the length of the trajectory). We then denote $\mathbf{a}$ as the identical initial condition for both teacher and student dynamics, and the superscript of relevant quantities (such as synaptic currents) is indexed by $\alpha$ or $\beta$. The trajectory distribution given the initial condition can be written in an explicit form following the Markovian property of the dynamics:
\begin{equation}\label{disE0}
\begin{aligned}
    p({\bx^{(\alpha)} }({t_1}),\ldots,{\bx^{(\alpha)} }({t_M})|\mathbf{a}) &= \prod\limits_l {p({\bx^{(\alpha)} }({t_l})|{\bx^{{(\alpha)}} }({t_{l - 1}}))}  \\
    &   = \prod\limits_{l,i} {\delta (x_i^{(\alpha )}({t_l}) -\blue{ F_i}[\bx^{(\alpha )}({t_{l - 1}})])} , 
\end{aligned}
\end{equation}
where ${\bx^{(\alpha)} }({t_0}) = \mathbf{a}$, and the function $F_i$ is defined below Eq.~\eqref{NSP}. We have discretized the original continuous dynamical equation into the following form:
\begin{equation}\label{disE}
x_i^{({\alpha})} ({t_l}) - x_i^{({\alpha})} ({t_{l - 1}}) =  - x_i^{({\alpha})} ({t_{l - 1}})h - \blue{\xi _i^{(\alpha)} ({t_{l - 1}})}h + \sum\limits_{j } {J_{ij}^{(\alpha)} } \phi _j^{(\alpha)} ({t_{l - 1}})h + {\delta _{l,0}}{a_i},
\end{equation}
where $\alpha=1,2$,  ${\delta _{l,0}}$ denotes the Kronecker symbol, and $\j^{(\alpha)}$ denotes an external current that will be set to zero at the final step.

Because our case is a deterministic dynamics, we can directly insert the $F$ function specified in Eq.~\eqref{disE} into the Dirac delta function in Eq.~\eqref{disE0}, and use further the Fourier representation of the delta function, and get the following result.
\begin{equation}
\footnotesize
\begin{aligned}
 &p({\bx^{(\alpha)} }({t_1}),\ldots,{\bx^{(\alpha)} }({t_M})|\mathbf{a},{\mathbf{J}^{(\alpha)} })\\
 &= \prod\limits_{l,i} {\delta (x_i^{({\alpha})} ({t_l}) - x_i^{({\alpha})} ({t_{l - 1}}) + x_i^{({\alpha})} ({t_{l - 1}})h + \blue{\xi _i^{(\alpha)} ({t_{l - 1}})}h - \sum\limits_{j } {J_{ij}^{(\alpha)} } {\phi _j}^{(\alpha)} ({t_{l - 1}})h - {\delta _{l,0}}{a_i})}    \\
 &  = \prod\limits_{l,i} {\int {\frac{{d{{\hat x}_i}^{(\alpha)} ({t_l})}}{{2\pi i}}\exp \left\{ {{{\hat x}_i}^{(\alpha)} ({t_l})\left[ {x_i^{({\alpha})} ({t_l}) - x_i^{({\alpha})} ({t_{l - 1}}) + x_i^{({\alpha})} ({t_{l - 1}})h + \blue{\xi _i^{(\alpha)} ({t_{l - 1}})}h - \sum\limits_{j } {J_{ij}^{(\alpha)} } \phi _j^{(\alpha)} ({t_{l - 1}})h - {\delta _{l,0}}{a_i}} \right]} \right\}} } , 
\end{aligned}
\end{equation}
where we have used $\delta (x) = \frac{1}{{2\pi i}}\int_{ - i\infty }^{i\infty } d \hat x{e^{\hat xx}}$ . Next, we  define the moment-generating functional
\begin{equation}
\footnotesize
\begin{aligned}
    &{Z^{(\alpha)} }(\bm{\xi}^{(\alpha)},\hat \j^{(\alpha)}|\a,{\J^{(\alpha)} })  = \prod\limits_{l,i} {\int {dx_i^{({\alpha})} (t)\exp \left( {{{\hat \xi}_i}^{(\alpha)} ({t_l})x_i^{({\alpha})} ({t_l})h} \right)} } p({x^{(\alpha)} }({t_1}),...,{x^{(\alpha)} }({t_M})|\a)\\
    & = \prod\limits_{l,i} {\int {\frac{{d{x_i}^{(\alpha)} ({t_l})d{{\hat x}_i}^{(\alpha)} ({t_l})}}{{2\pi i}}} } \exp \left( {{{\hat x}_i}^{(\alpha)} ({t_l})\left[ {x_i^{({\alpha})} ({t_l}) - x_i^{({\alpha})} ({t_{l - 1}}) + x_i^{({\alpha})} ({t_{l - 1}})h - \sum\limits_{j } {J_{ij}^{(\alpha)} } \phi _j^{(\alpha)} ({t_{l - 1}})h - {\delta _{l,0}}{a_i}} \right]} \right)\\
    &    \times \exp \left({{\hat \xi}_i}^{(\alpha)} ({t_l})x_i^{({\alpha})} ({t_l})h + \blue{\xi_i^{(\alpha)} ({t_{l - 1}})}{{\hat x}_i}^{(\alpha)} ({t_l})h\right),\\
      &=\int \prod\limits_{l,i} \frac{dx_i^{({\alpha})} (t_l)d{\hat x}_i^{(\alpha)} (t_l)}{{2\pi i}} \exp \left(h\sum\limits_{li} {\hat x}_i^{(\alpha)} ({t_l}) \left[\frac{{x_i^{({\alpha})} ({t_l}) - x_i^{({\alpha})} ({t_{l - 1}})}}{h} + x_i^{({\alpha})} ({t_{l - 1}}) - \sum\limits_{j } {J_{ij}^{(\alpha)} } \phi _j^{(\alpha)} ({t_{l - 1}}) - \frac{{{\delta _{l,0}}{a_i}}}{h}\right]\right)\\
      &\times \exp\left(\sum_{il}{{\hat \xi}_i}^{(\alpha)} ({t_l})x_i^{({\alpha})} ({t_l})h + \blue{\xi_i^{(\alpha)} ({t_{l - 1}})}{{\hat x}_i}^{(\alpha)} ({t_l})h\right).
\end{aligned}
\end{equation}

Now, we take the continuous limit as $M \rightarrow \infty$ , $h \rightarrow 0$ yet $Mh = T$, and obtain $h\sum\limits_{l = 0}^M f ({t_l}) = \int f (t){\rm{d}}t$. Therefore, the limit $\mathop {\lim }\limits_{h \to 0} \frac{{{x_i}({t_l}) - {x_i}({t_{l - 1}})}}{h} = {{\dot x}_i}$. We then introduce the shorthand $\prod_{i, l} \mathrm{~d} x_i^\alpha\left(t_l\right) \xrightarrow{h \rightarrow 0} D \mathbf{x}^\alpha$ and $\prod_{i, l} \frac{\mathrm{d} \hat{x}_i^\alpha\left(t_l\right)}{2 \pi i} \xrightarrow{h \rightarrow 0} D \hat{\mathbf{x}}^\alpha$. Under this continuous limit, the moment generating functional reads:
\begin{equation}
\begin{aligned}
Z^{(\alpha)}\left(\bm{\xi}^{(\alpha)}, \hat{\j}^{(\alpha)} \mid \a, \J^{(\alpha)}\right)= & \int D \mathbf{x}^{(\alpha)} D \hat{\mathbf{x}}^{(\alpha)} \exp \left(\sum_i\left[\int_0^T \hat{x}_i^{(\alpha)}(t)\left[\dot{x}_i^{(\alpha)}(t)+x_i^{({\alpha})}(t)-a_i \delta(t)\right] d t\right]\right) \\
& \times \exp \left(\sum_i \int_0^T\left[\hat{\xi}_i^{(\alpha)}(t) x_i^{({\alpha})}(t)+\blue{\xi_i^{(\alpha)}(t)} \hat{x}_i^{(\alpha)}(t)\right] d t\right) \\
& \times \exp \left(-\sum_i \int_0^T \hat{x}_i^{(\alpha)}(t)\left[\sum_{j} J_{i j}^{(\alpha)} \phi_j^{(\alpha)}(t)\right] d t\right).
\end{aligned}
\end{equation}

The moment-generating function for the combined system of two networks reads
\begin{equation}\label{GF}
    Z = \prod\limits_{{\alpha} } {{Z^{(\alpha)} }({\bm{\xi}^{(\alpha)} },{{\hat \j}^{(\alpha)} }|\a,{\J^{(\alpha)} })}.
\end{equation}
Given the respective coupling and the same initial condition, both network dynamics evolve independently.
Next, we perform the disorder average over $\{J_{ij}^{(\alpha)}\}$. 
\blue{
\begin{equation}\label{ave}
\begin{aligned}
& \left\langle\exp \left(-\sum_{i\neq j, {\alpha}} \int_0^T \hat{x}_i^{(\alpha)}(t)\left[J_{i j}^{(\alpha)} \phi_j^{(\alpha)}(t)\right] d t\right)\right\rangle_{\J^{(1)},\J^{(2)}} \\
& \simeq\prod_{i,j}\left\langle\exp \left(-\sum_{\alpha} \int_0^T \hat{x}_i^{(\alpha)}(t)\left[J_{i j}^{(\alpha)} \phi_j^{(\alpha)}(t)\right] d t\right)\right\rangle \\
&= \prod\limits_{i , j} {\left[ {\exp\left( {\frac{{{g^2}}}{{2N}}\int_0^T\int_0^T {\hat x_i^{(1)}} (s)\hat x_i^{(1)}(t)\phi _j^{(1)}(s)\phi _j^{(1)}(t)dsdt + \frac{{{g^2}}}{{2N}}\int_0^T\int_0^T {\hat x_i^{(2)}} (s)\hat x_i^{(2)}(t)\phi _j^{(2)}(s)\phi _j^{(2)}(t)dsdt} \right.} \right.}   \\
& \left. {\left. { + \frac{{{g^2}\sqrt {1 - {\eta ^2}} }}{N}\int_0^T\int_0^T {\hat x_i^{(1)}} (s)\hat x_i^{(2)}(t)\phi _j^{(1)}(s)\phi _j^{(2)}(t)dsdt} \right)} \right].
\end{aligned}
\end{equation}
}
The appearance of the cross term in the exponential is due to our perturbation scheme $J_{ij}^{(2)} = \sqrt {1 - {\eta ^2}} J_{ij}^{(1)} + \eta \delta J_{ij}^{(1)}$. 
We also add the negligible diagonal terms of the coupling matrix in Eq.~\eqref{ave}, similar to that in Ref.~\cite{Qiu-2024}.

By an inspection of the last line of Eq.~\eqref{ave}, we have to introduce the dynamics order parameters, i.e., correlation functions:
\begin{subequations}
\begin{align}
   {T_1}(s,t) &= \frac{{{g^2}}}{N}\sum\limits_j {\phi _j^{(1)}} \left( s \right)\phi _j^{(2)}\left( t \right),\\
  Q_1^{(\alpha)} (s,t) &= \frac{{{g^2}}}{N}\sum\limits_j {\phi _j^{(\alpha)} } \left( s \right)\phi _j^{(\alpha)} \left( t \right).
\end{align}
\end{subequations}
From this step, one can see that the apparent order emerges from the intrinsic disorder in the coupling statistics, which is similar to what occurs in replica calculation of equilibrium models~\cite{Huang-2022} and non-equilibrium steady states~\cite{Qiu-2024}. Then we can insert the definition of the order parameters
by introducing an integral of Dirac delta functions together with the Fourier representation of the delta function, leading to the following result:
\begin{subequations}
\begin{align}
    &\prod\limits_{s,t} {\delta \left( { - \frac{N}{{{g^2}}}Q_1^{(\alpha)} \left( {s,t} \right) + \sum\limits_j {\phi _j^{(\alpha)} } \left( s \right)\phi _j^{(\alpha)} \left( t \right)} \right)}\notag \\
    &  = \int D Q_2^{(\alpha)} \left( {s,t} \right)\exp \left[ {\int_0^T\int_0^T {Q_2^{(\alpha)} \left( {s,t} \right)} \left( { - \frac{N}{{{g^2}}}Q_1^{(\alpha)} \left( {s,t} \right) + \sum\limits_j {\phi _j^{(\alpha)} } \left( s \right)\phi _j^{(\alpha)} \left( t \right)} \right){\rm{d}}s{\rm{d}}t} \right],\\
   &\prod\limits_{s,t} {\delta \left( { - \frac{N}{{{g^2}}}{T_1}(s,t) + \sum\limits_j {\phi _j^{(1)}} \left( s \right)\phi _j^{(2)}\left( t \right)} \right)}\notag \\
   &  = \int D {T_2}(s,t)\exp \left[ {\int_0^T\int_0^T {{T_2}(s,t)} \left( { - \frac{N}{{{g^2}}}{T_1}(s,t) + \sum\limits_j {\phi _j^{(1)}} \left( s \right)\phi _j^{(2)}\left( t \right)} \right){\rm{d}}s{\rm{d}}t} \right],
\end{align}
\end{subequations}
where \blue{$DX_2\equiv\prod_{s,t}\frac{1}{2\pi i}dX_2$} in which $X_2=Q_2^{(\alpha)}$, or $T_2$. The purely imaginary field $X_2\in i\mathbb{R}$.

By collecting all terms into Eq.~\eqref{GF}, we obtain a compact expression of the moment generating functional:
\blue{
\begin{equation}
\footnotesize
\begin{aligned}
       Z& = \prod\limits_{{\alpha}} {{Z^{(\alpha)} }({\bm{\xi}^{(\alpha)} },{{\hat \j}^{(\alpha)} }|\a)} \\& = \prod\limits_{\alpha}  {\left\{ {\int D Q_1^{(\alpha)} \int D Q_2^{(\alpha)} \int D {{\bf{x}}^{(\alpha)} }D{{\hat {\bf{x}}}^{(\alpha)} }} \right\}} \int D {T_1}D{T_2}\exp \left( {\sum\limits_{i,\alpha} {\left[ {\int_0^T {\hat x_i^{({\alpha})} } (t)\left[ {\dot x_i^{({\alpha})} (t) + x_i^{({\alpha})} (t) - {a_i}\delta (t)} \right]dt} \right]} } \right)\\
      & \times \exp \left(\sum_{i,\alpha} \int_0^T\left[\hat{\xi}_i^{(\alpha)}(t) x_i^{({\alpha})}(t)+\xi_i^{(\alpha)}(t) \hat{x}_i^{(\alpha)}(t)\right] d t\right) \\
&  \times \prod\limits_i  \exp\left( {\frac{1}{2}\int_0^T\int_0^T {\hat x_i^{(1)}} (s)\hat x_i^{(1)}(t)Q_1^{(1)}(s,t)dsdt + \frac{1}{2}\int_0^T \int_0^T{\hat x_i^{(2)}} (s)\hat x_i^{(2)}(t)Q_1^{(2)}(s,t)dsdt} \right.   \\
& \left. { + \sqrt {1 - {\eta ^2}} \int_0^T\int_0^T {\hat x_i^{(1)}} (s)\hat x_i^{(2)}(t){T_1}(s,t)dsdt} \right)\\
& \times \prod\limits_{\alpha}  {\exp \left[ {\int_0^T\int_0^T {Q_2^{(\alpha)} \left( {s,t} \right)} \left( { - \frac{N}{{{g^2}}}Q_1^{(\alpha)} \left( {s,t} \right) + \sum\limits_j {\phi _j^{(\alpha)} } \left( s \right)\phi _j^{(\alpha)} \left( t \right)} \right){\rm{d}}s{\rm{d}}t} \right]} \\
& \times\exp \left[ {\int_0^T\int_0^T {{T_2}(s,t)} \left( { - \frac{N}{{{g^2}}}{T_1}(s,t) + \sum\limits_j {\phi _j^{(1)}} \left( s \right)\phi _j^{(2)}\left( t \right)} \right){\rm{d}}s{\rm{d}}t} \right],
\\&=\prod\limits_{\alpha}  {\left\{ {\int D Q_1^{(\alpha)} \int D Q_2^{(\alpha)} } \right\}} \int\int D {T_1}D{T_2} {\exp \left(\Omega \left[ {{{\left\{ {Q_1^{(\alpha)} ,Q_2^{(\alpha)} } \right\}}_{{\alpha}  \in \{ 1,2\} }},{T_1},{T_2}} \right]\right)},
\end{aligned}
\end{equation}}
where the dynamical action reads
\begin{align}
\Omega \left[ {{{\left\{ {Q_1^{(\alpha)} ,Q_2^{(\alpha)} } \right\}}_{{\alpha}  \in \{ 1,2\} }},{T_1},{T_2}} \right] =  - \frac{N}{{{g^2}}}\int_0^T\int_0^T {{T_2}{T_1}dsdt}  - \sum\limits_{\alpha}  {\frac{N}{{{g^2}}}} \int_0^T\int_0^T {Q_2^{(\alpha)} Q_1^{(\alpha)} dsdt}  + \ln Z'.
\end{align}
The auxiliary quantity $Z'$ is given below:
\begin{equation}\label{Zprim}
\begin{aligned}
Z^{\prime} &    = \int {\prod\limits_{\alpha}  {D{{\bf{x}}^{(\alpha)} }D{{{\bf{\hat x}}}^{(\alpha)} }} }  \exp \left( {\sum\limits_{\alpha ,i} {\left[ {\int_0^T {\hat x_i^{({\alpha})} } (t)\left[ {\dot x_i^{({\alpha})} (t) + x_i^{({\alpha})} (t) - {a_i}\delta (t)} \right]dt} \right]} } \right) \\
&  \times \exp \left( {\sum\limits_{{\alpha} ,i} {\int_0^T {\left[ {\hat {\xi}_i^{(\alpha)} (t)x_i^{({\alpha})} (t) + \blue{{\xi_i}^{(\alpha)} (t)}\hat x_i^{({\alpha})} (t)} \right]} } dt} \right) \\
&   \times \exp \left( {\frac{1}{2}\sum\limits_{{\alpha} ,i} {\int_0^T\int_0^T {\hat x_i^{({\alpha})} } } (s)\hat x_i^{({\alpha})} (t)Q_1^{(\alpha)} dsdt + \int_0^T\int_0^T {\sum\limits_{{\alpha} ,j} {\phi _j^{(\alpha)} } } (s)\phi _j^{(\alpha)} (t)Q_2^{(\alpha)} {\rm{d}}s\;{\rm{d}}t} \right)\\
&  \times \exp \left( {\sqrt {1 - {\eta ^2}} \sum\limits_i {\int_0^T\int_0^T {\hat x_i^{(1)}} } (s)\hat x_i^{(2)}(t){T_1}dsdt + \int_0^T \int_0^T{\sum\limits_j {\phi _j^{(1)}} } (s)\phi _j^{(2)}(t){T_2}\;{\rm{d}}s\;{\rm{d}}t} \right).
\end{aligned}
\end{equation}
A careful inspection of Eq.~\eqref{Zprim} reveals the factorization over the neuron index $i$, and we also find that each term in the factorization bears the same form. Therefore, one can reduce the auxiliary quantity to a low-dimensional integral or one-dimensional mean-field dynamics for the combined system as follows,
\begin{equation}
       Z = \prod\limits_{\alpha}  {\left\{ {\int D Q_1^{(\alpha)} \int D Q_2^{(\alpha)} } \right\}} \int D {T_1}D{T_2}\exp \left( {Nf\left[ {{{\left\{ {Q_1^{(\alpha)} ,Q_2^{(\alpha)} } \right\}}_{{\alpha}  \in \{ 1,2\} }},{T_1},{T_2}} \right]} \right),
\end{equation}
where we obtain a dynamical free-entropy $f\left[ \left\{ {Q_1^{(\alpha)} ,Q_2^{(\alpha)} } \right\}_{\alpha  \in \{ 1,2\} },{T_1},{T_2} \right]$:
\begin{equation}
 f =  - \frac{1}{{{g^2}}}\int_0^T\int_0^T {{T_2}{T_1}dsdt}  - \sum\limits_{\alpha}  {\frac{1}{{{g^2}}}} \int_0^T\int_0^T {Q_2^{(\alpha)} Q_1^{(\alpha)} dsdt}  + \ln {Z^{12}}.
\end{equation}
The effective generating functional for the combined system $Z^{12}$ can be compactly written as
\begin{equation}
\begin{aligned}
    {Z^{12}} & = \int {\prod\limits_{\alpha}  {D{{\bf{x}}^{(\alpha)} }D{{{\bf{\hat x}}}^{(\alpha)} }} }  \exp \left( {\sum\limits_{\alpha}  {\left[ {\int\limits_0^T {{{\hat x}^{(\alpha)} }(t)[{{\dot x}^{(\alpha)} }(t)}  + {x^{(\alpha)} }(t) - a\delta (t)]dt} \right]} } \right)\\
    &  \times \exp \left( {\sum\limits_{\alpha}  {\int\limits_0^T {[{{\hat \xi}^{(\alpha)} }(t){x^{(\alpha)} }(t) + \blue{{\xi^{(\alpha)} }(t)}{{\hat x}^{(\alpha)} }(t)]dt} } } \right)\\
    &  \times \exp \left( {\sum\limits_{\alpha}  {\left[ {\frac{1}{2}\int_0^T\int_0^T {{{\hat x}^{(\alpha)} }(s){{\hat x}^{(\alpha)} }(t)Q_1^{(\alpha)} } dsdt + \int_0^T\int_0^T {{\phi ^{(\alpha)} }\left( s \right){\phi ^{(\alpha)} }\left( t \right)Q_2^{(\alpha)} {\rm{d}}s{\rm{d}}t} } \right]} } \right)\\
    &  \times \exp \left( {\sqrt {1 - {\eta ^2}} \int_0^T\int_0^T {{{\hat x}^{(1)}}(s){{\hat x}^{(2)}}(t){T_1}} dsdt + \int_0^T\int_0^T {{\phi ^{(1)}}\left( s \right){\phi ^{(2)}}\left( t \right){T_2}{\rm{d}}s{\rm{d}}t} } \right).
\end{aligned}
\end{equation}

In the thermodynamic limit, the free-entropy must take a maximum that yields $\ln Z=Nf^*$, where $*$ means that all relevant order parameters must take specific values making the dynamical free-entropy maximal. This step is also called the saddle point approximation or Laplace method. In detail, we next calculate the following zero-gradient equations:
\begin{equation}
\begin{aligned}
& \frac{\delta f}{\delta Q_2^{(\alpha)}}=0 \rightarrow Q_1^{{(\alpha)} *}=g^2\left\langle\phi^{(\alpha)}(s) \phi^{(\alpha)}(t)\right\rangle, \\
&\blue{\frac{{\delta f}}{{\delta Q_1^{(\alpha)} }} = 0 \to Q_2^{{{(\alpha)} ^*}} = \frac{{{g^2}}}{2}\left\langle {{{\hat x}^{(\alpha)} }(s){{\hat x}^{(\alpha)} }(t)} \right\rangle  = 0,}  \\
& \frac{\delta f}{\delta T_2}=0 \rightarrow T_1^*=g^2\left\langle\phi^{(1)}(s) \phi^{(2)}(t)\right\rangle, \\
& \frac{\delta f}{\delta T_1}=0 \rightarrow T_2^*=g^2 \sqrt{1-\eta^2}\left\langle\hat{x}^{(1)}(s) \hat{x}^{(2)}(t)\right\rangle=0,
\end{aligned}
\end{equation}
where the average $\langle\cdot\rangle$ is performed under the measure given by the effective partition function $Z^{12}$~\cite{Zou-2024}. \blue{More precisely, we write 
$Z^{12}=\int\mathcal{D}X e^{\mathcal{L}(X)}$, where $X$ indicates all relevant quantities whose integral measure is defined by $\mathcal{D}X$, and then the average of any observable $\mathcal{O}(X)$ is expressed as $\langle\mathcal{O}(X)\rangle=\frac{\int\mathcal{D}X\mathcal{O}(X)e^{\mathcal{L}(X)}}{Z^{12}}$.}
 We remark that all expectations of only $\hat{x}^\alpha$ vanishes due to the normalization of trajectory probability, i.e., $Z(\blue{\bm\xi},\hat{\j}=0)=1$, \blue{and $\frac{{{\delta ^n}}}{{\delta {\xi ^{(\alpha)} }\left( {{t_1}} \right) \cdots \delta {\xi ^{(\beta)} }\left( {{t_n}} \right)}}Z(\blue{\bm\xi},\hat{\j}=0) = \left\langle {{{\hat x}^{(\alpha)} }\left( {{t_1}} \right) \cdots {{\hat x}^{(\beta)} }\left( {{t_n}} \right)} \right\rangle  \equiv 0$}~\cite{Zou-2024,Helias-2020}.

To conclude, at the maximum, the dynamical partition function $Z=(Z^*)^N$, where $Z^*$ is given below:
\begin{equation}\label{epath}
\begin{aligned}
    {Z^ * }&= \int {\prod\limits_{\alpha}  {D{{\bf{x}}^{(\alpha)} }D{{{\bf{\hat x}}}^{(\alpha)} }} }  \exp \left( {\sum\limits_{\alpha}  {\left[ {\int\limits_0^T {{{\hat x}^{(\alpha)} }(t)[{{\dot x}^{(\alpha)} }(t)}  + {x^{(\alpha)} }(t) - a\delta (t)]dt} \right]} } \right)\\
    &  \times \exp \left( {\sum\limits_{\alpha}  {\int\limits_0^T {[{{\hat \xi}^{(\alpha)} }(t){x^{(\alpha)} }(t) +\blue{ {\xi^{(\alpha)} }(t)}{{\hat x}^{(\alpha)} }(t)]dt} } } \right)\\
    &   \times \exp \left( {\sum\limits_{\alpha}  {\left[ {\frac{1}{2}\int_0^T\int_0^T {{{\hat x}^{(\alpha)} }(s){{\hat x}^{(\alpha)} }(t)Q_1^{{(\alpha)} *}} dsdt} \right]} } \right)\\
    &     \times \exp \left( {\sqrt {1 - {\eta ^2}} \int_0^T\int_0^T {{{\hat x}^{(1)}}(s){{\hat x}^{(2)}}(t)T_1^*} dsdt} \right).
\end{aligned}
\end{equation}
The effective two-variable dynamical partition function suggests that the high-dimensional dynamics of combined systems is in principle able to be captured by the following two-dimensional mean-field dynamics:
\begin{equation}\label{emf}
    \left(\partial_t+1\right) x^{(\alpha)}(t)=\gamma^{(\alpha)}(t) \quad {(\alpha)} \in\{1,2\},
\end{equation}
where the zero-mean effective noise (despite the deterministic nature of the dynamics) appears from the above calculation as follows,
\begin{subequations}\label{cov}
\begin{align}
   &\left\langle {{\gamma ^{(\alpha)} }(s){\gamma ^{(\beta)} }(t)} \right\rangle  = {g^2}\left\langle {{\phi ^{(\alpha)} }(s){\phi ^{(\beta)} }(t)} \right\rangle \left[(1 - {\delta _{{\alpha} {\beta} }})\sqrt {1 - {\eta ^2}}  + {\delta _{{\alpha} {\beta} }}\right],
    \\&    {\phi ^{(\alpha)} }(s) = \tanh [{x^{(\alpha)} }(s)],\\
    &{x^{(1)}}(0) = {x^{(2)}}(0) = a,\\&
    a \sim \mathcal{N}\left(0, 1\right).
\end{align}
\end{subequations}
\blue{Equation~\eqref{emf} is consistent with Eq.~\eqref{epath} in a path integral representation of the stochastic dynamics. To show this, we first calculate the following expectation:
\begin{equation}\label{cov2}
    \begin{aligned}
        &\left\langle {\exp\left (\sum\limits_\alpha  {\int\limits_0^T {{{\hat x}^{(\alpha )}}(t){\gamma ^{(\alpha )}}(t)} } dt\right)} \right\rangle  = \exp \left(\frac{1}{2}{\left\langle {\left[\sum\limits_\alpha  {\int\limits_0^T {{{\hat x}^{(\alpha )}}(t){\gamma ^{(\alpha )}}(t)} } dt\right]^2} \right\rangle }\right)\\& = \exp\left (\frac{1}{2}\sum\limits_\alpha  {\int\limits_0^T \int\limits_0^T{\left\langle {{\gamma ^{(\alpha )}}(t){\gamma ^{(\alpha )}}(s)} \right\rangle } {{\hat x}^{(\alpha )}}(t)} {{\hat x}^{(\alpha )}}(s)dtds + \int\limits_0^T \int\limits_0^T{\left\langle {{\gamma ^{(1)}}(t){\gamma ^{(2)}}(s)} \right\rangle } {{\hat x}^{(1)}}{{\hat x}^{(2)}}(s)dtds\right),
    \end{aligned}
\end{equation}
where the first equality can be proved by using the Cholesky decomposition of
the covariance (or a multivariate Hubbard-Stratonovich transformation). By comparing Eq.~\eqref{cov} and Eq.~\eqref{cov2}, one arrives at the following statistics for the colored noise $\gamma^{(\alpha)}(t)$:
\begin{subequations}
\begin{align}
\left\langle {{\gamma ^{(\alpha )}}(t){\gamma ^{(\alpha )}}(s)} \right\rangle &= Q_1^{(\alpha )*} = {g^2}\left\langle {{\phi ^{(\alpha )}}(s){\phi ^{(\alpha )}}(t)} \right\rangle, \\
\left\langle {{\gamma ^{(1)}}(t){\gamma ^{(2)}}(s)} \right\rangle & = \sqrt {1 - {\eta ^2}} T_1^* = \sqrt {1 - {\eta ^2}} {g^2}\left\langle {{\phi ^{(1)}}(s){\phi ^{(2)}}(t)} \right\rangle.
\end{align}
\end{subequations}
}

\section{Numerical procedure of solving the dynamical mean-field dynamics}\label{app-c}
In this section, we detail the numerical procedure to solve the DMFT equations in our theory. 
We set $L$ to be the total running time, and $h$ represents the discrete time step size. At the beginning of the iteration, we sample $M$ points from a standard Gaussian distribution as the common initial points of ${\bx^{(1)}}(0)$ and ${\bx^{(2)}}(0)$, satisfying ${\bx^{(1)}}(0) = {\bx^{(2)}}(0)$. Then we initialize the effective random noise vector ${{\gamma ^{(\alpha)} }(t)}$ with the mean zero and the covariance function matrix $\left\langle {{\gamma ^{(\alpha)} }(s){\gamma ^{(\beta)} }(t)} \right\rangle $ [see Eq.~\eqref{cov}], whose dimensions are $\left( {2L/h} \right) \times \left( {2L/h} \right)$ for $sh\le L$ and $th\le L$, and a random guess is assigned to the covariance matrix at the initial step. In each round of iteration, we carry out the following steps:
\begin{enumerate}
    \item Draw $M$ samples of noise trajectories $\left\{ {\left\{ {{\gamma_m^{(\alpha)} }(t)} \right\}_{{(\alpha)}  = 1}^2} \right\}_{m = 1}^M$ from the multivariate Gaussian distribution $\mathcal{N}\left(0, \left\langle {{\gamma ^{(\alpha)} }(t){\gamma ^{(\beta)} }(s)} \right\rangle \right)$. This is can be achieved by a Cholesky decomposition of the covariance and then using the lower-triangular matrix of the decomposition to generate the required samples~\cite{Zhou-2021}.
    \item For these noise trajectories, run $M$ corresponding current trajectories independently by using the following direct discretization:
$$
{x^{(\alpha)}}(t + h) = (1 - h){x^{(\alpha)}}(t) + h{\gamma ^{(\alpha)} }(t) \quad {(\alpha)} \in\{1,2\}.
$$
    \item  The noise covariance function $\left\langle {{\gamma ^{(\alpha)} }(t){\gamma ^{(\beta)} }(s)} \right\rangle $ can be refreshed by 
\[\left\langle {{\gamma ^{(\alpha)} }(s){\gamma ^{(\beta)} }(t)} \right\rangle  = {g^2}\left\langle {{\phi ^{(\alpha)} }(s){\phi ^{(\beta)} }(t)} \right\rangle \left[(1 - {\delta _{{\alpha} {\beta} }})\sqrt {1 - {\eta ^2}}  + {\delta _{{\alpha} {\beta} }}\right],\]
where $\left\langle {{\phi ^{(\alpha)} }(s){\phi ^{(\beta)} }(t)} \right\rangle $ represents the mean over $M$ corresponding current trajectories.
More precisely, $$\left\langle {{\phi ^{(\alpha)} }(s){\phi ^{(\beta)} }(t)} \right\rangle  = \frac{1}{M}\sum\limits_{m = 1}^M \tanh [x^{(\alpha)}_m(s)]\tanh [x_m^{(\beta)} (t)]. $$
\end{enumerate}
In the third step, the new covariance matrix is updated by using a damping trick to speed up convergence.

Finally,  the evolution of mean square distance can also be estimated by using the ensemble of neural states generated during the above numerical procedure.
\begin{equation}
    d(t) = \left\langle {{x^{(1)}}(t){x^{(1)}}(t)} \right\rangle  + \left\langle {{x^{(2)}}(t){x^{(2)}}(t)} \right\rangle  - 2\left\langle {{x^{(1)}}(t){x^{(2)}}(t)} \right\rangle .
\end{equation}

\section{Numerical procedure of estimating the prediction Lyapunov exponent}\label{app-d}
By analogy to the standard Lyapunov exponent estimation, we adopt the orbit separation method to compute the prediction Lyapunov exponent. Readers can refer to the book~\cite{JC-2003} for details of the standard numerical calculation of the Lyapunov exponent characterizing the sensitivity of the dynamics to initial uncertainty. Here,  we outline the detailed procedure below, where we focus on the deviation of two obits starting from the same initial condition but with slightly different model parameters.

The procedure is described as follows.
\begin{enumerate}
    \item We sample an initial point from an 
$N$-dimensional standard Gaussian distribution, and then generate a first step of a trajectory $\bx^{(1)}$ with unperturbed weights, and the same for the other trajectory $\bx^{(2)}$ of the perturbed system. \blue{Note that both trajectories share the same initial point.}
The initial state-deviation can be calculated $\epsilon=\|\bx^{(1)}(t=1)-\bx^{(2)}(t=1)\|_2^2$. $t$ means discrete time step, and the real time is obtained by $ht$.
    \item Run the dynamics forward in one step, and get $\boldsymbol\Delta(t=2)=\bx^{(1)}(t=2)-\bx^{(2)}(t=2)$. The first exponent is calculated as $\lambda_1=\ln\left(\frac{\|\boldsymbol\Delta(t=2)\|_2}{\epsilon^{1/2}}\right)$.
    \item  Apply a renormalization step to ensure the orbits are separated by a magnitude of $\epsilon$, maintaining the direction of maximal expansion:
    $$\bx^{(2)}(t)=\bx^{(1)}(t)+\epsilon^{1/2}\frac{\boldsymbol\Delta(t)}{\|\boldsymbol\Delta(t)\|_2}.$$
    \item Iterate the above procedure many times for each of them a Lyapunov exponent is estimated in an analog way to $\lambda_1$.  The second trajectory is then renormalized before running the second dynamics in an additional step.
    \item We finally compute the average of the different separation rates as the Lyapunov exponent for the dynamics prediction:
    $$\lambda_{\rm pd}  = \frac{1}{(n-t_0+1)h}\sum\limits_{s=t_0}^n {{\lambda _s}} .$$
    Note that the above sum refers to the steady regime, i.e., the sum starts from $t_0$.
\end{enumerate}

In our simulations, we set the time increment $h=0.01$, the number of neurons $N = 1\,000$, and the weight deviation $\eta  = 0.01$. For different values of $\eta$, we found that 
$\epsilon^{1/2}$ is a linear function of $g$ in our numerical simulation. When $\eta\to 0$, $\epsilon$ is very small as well. For comparison, we also calculate the standard Lyapunov exponent whose behavior is shown in the upper inset of Fig.~\ref{fig9}. The procedure is similar but the initial separation is 
randomly chosen, such that $\bx^{(2)}=\bx^{(1)}+\epsilon\frac{\boldsymbol\Delta(0)}{\|\boldsymbol\Delta(0)\|_2}$, where $\boldsymbol\Delta(0)$ is a random vector of dimension $N$, and $\epsilon$ is a pre-defined small quantity. 



\end{document}